\documentclass[12pt]{article}

\usepackage{tikz, adjustbox, float} 
\usepackage[round]{natbib} 
\usepackage{amsmath, amsfonts, amssymb, amsthm, dsfont,mathabx} 
\usepackage{mathrsfs} 
\usepackage{soul} 
\usepackage{bm} 
\usepackage{booktabs}  
\usepackage{enumitem} 
\usepackage[utf8]{inputenc}
\usepackage{titlesec}
\titlespacing{\section}{0pt}{\parskip}{-\parskip}
\titlespacing{\subsection}{0pt}{\parskip}{-\parskip}
\titlespacing{\subsubsection}{0pt}{\parskip}{-\parskip}

\def\real{{\rm I\!R}}

\DeclareMathOperator*{\argmin}{argmin}

\DeclareMathOperator*{\Int}{Int}
\DeclareMathOperator*{\cov}{cov}
\def\0{{\bf 0}}
\def\1{{\bf 1}}

\def\bpi{{\bm{\pi}}}
\def\hbpi{\wh{{\bm{\pi}}}}

\def\bt{{\boldsymbol\theta}}
\def\bT{{\boldsymbol\Theta}}
\def\bb{{\boldsymbol\beta}}

\def\bomega{\boldsymbol\omega}

\def\bOmega{{\bm{\Omega}}}

\def\x{{\bf x}}
\def\I{{\bf I}}
\def\Y{\mathbf{Y}}
\def\y{{\bf y}}

\DeclareMathOperator*{\argzero}{argzero}

\usepackage{latexsym}
\usepackage{amsthm}
\usepackage{amsmath,amssymb,amsfonts,bm,amsbsy,bbm}
\usepackage{epsfig}
\usepackage{graphicx,rotating,lscape}
\usepackage[colorlinks=true,  urlcolor=blue, citecolor=blue, psdextra, pdfencoding=auto, hypertexnames=false]{hyperref}
\usepackage{verbatim}
\usepackage{natbib}
\usepackage{multirow,color}
\usepackage{geometry}
\usepackage{setspace}
\usepackage{subfigure}
\usepackage{bbm}
\usepackage{ulem} 
\usepackage{mathtools}

\usepackage[capitalise]{cleveref}

\def\x{{\bf x}}

\def\Z{{\bf Z}}

\def\0{{\bf 0}}

\def\wh{\widehat}

\def\wc{\widecheck}

\def\cov{\hbox{cov}}

\def\bse{\begin{eqnarray*}}
\def\ese{\end{eqnarray*}}
\def\be{\begin{eqnarray}}
\def\ee{\end{eqnarray}}
\def\bsq{\begin{equation*}}
\def\esq{\end{equation*}}
\def\bq{\begin{equation}}
\def\eq{\end{equation}}

\def\boxit#1{\vbox{\hrule\hbox{\vrule\kern6pt  \vbox{\kern6pt#1\kern6pt}\kern6pt\vrule}\hrule}}
\def\bse{\begin{eqnarray*}}
\def\ese{\end{eqnarray*}}
\def\be{\begin{eqnarray}}
\def\ee{\end{eqnarray}}
\def\bsq{\begin{equation*}}
\def\esq{\end{equation*}}
\def\bq{\begin{equation}}
\def\eq{\end{equation}}

\def\cov{\hbox{cov}}

\def\wh{\widehat}

\def\cov{\mbox{cov}}

\def\d{{\bf d}}

\def\I{{\bf I}}

\def\W{{\bf W}}
\def\w{{\bf w}}
\def\X{{\bf X}}

\def\x{{\bf x}}

\def\Y{{\bf Y}}

\def\y{{\bf y}}
\def\Z{{\bf Z}}

\def\bSig{{\bf \Sigma}}

\def\dovr{\buildrel d\over\longrightarrow}

\def\log{\hbox{log}}

\def\squarebox#1{\hbox to #1{\hfill\vbox to #1{\vfill}}}

\def\bpi{{\boldsymbol \pi}}
\def\0{{\bf 0}}

\def\bcalW{\bm{\mathcal{W}}}

\def\cov{\hbox{cov}}

\def\wh{\widehat}

\def\log{\hbox{log}}

\usepackage{nicefrac}

\newtheoremstyle{mytheoremstyle} 
    {0.3cm}                      
    {0cm}                        
    {\it}                   
    {}                           
    {\bf}                   
    {: }                          
    {0em}                       
    {}  

\theoremstyle{mytheoremstyle}
\newtheorem{Theorem}{Theorem}

\newtheorem*{Lemma*}{Lemma}

\newtheoremstyle{myExampleRemarkstyle} 
    {0.3cm}                    
    {0cm}                      
    {\it}                         
    {}                         
    {\bf}                      
    {: }                       
    {0em}                      
    {}  

\theoremstyle{myExampleRemarkstyle}
\newtheorem{Example}{Example} 
\newtheorem{Remark}{Remark}
\newtheorem{Assumption}{Assumption}

\newtheorem{innercustomgeneric}{\customgenericname}
\providecommand{\customgenericname}{}
\newcommand{\newcustomtheorem}[2]{%
  \newenvironment{#1}[1]
  {%
   \renewcommand\customgenericname{#2}%
   \renewcommand\theinnercustomgeneric{##1}%
   \innercustomgeneric
  }
  {\endinnercustomgeneric}
}

\newcustomtheorem{customthm}{Theorem}
\newcustomtheorem{customlemma}{Lemma}
\usepackage{mathalfa}
\usepackage{oubraces}

\let\refBKP\ref
\renewcommand{\ref}[1]{{\upshape\refBKP{#1}}}

\def\boxit#1{\vbox{\hrule\hbox{\vrule\kern3pt
          \vbox{\kern3pt#1\kern3pt}\kern3pt\vrule}\hrule}}

\usepackage{shellesc}
\ShellEscape{cp /usr/local/share/latexmk/LatexMk ./LatexMk}

\begin{document}

\begin{center}
    {\LARGE An accurate percentile method for parametric inference based on asymptotically biased estimators}  \\
   \vspace{1cm}
    Samuel Orso$^\ast$, Mucyo Karemera$^\ast$,
    Maria-Pia Victoria-Feser \&
    Stéphane Guerrier\\

    $^\ast$ The first two authors contributed equally.
\end{center}

\begin{abstract}
Inference methods for computing confidence intervals in parametric settings usually rely on consistent estimators of the parameter of interest. However, it may be computationally and/or analytically burdensome to obtain such estimators in various parametric settings, for example when the data exhibit certain features such as censoring, misclassification errors or outliers. To address these challenges, we propose a simulation-based inferential method, called the implicit bootstrap, that remains valid regardless of the potential asymptotic bias of the estimator on which the method is based. We demonstrate that this method allows for the construction of asymptotically valid percentile confidence intervals of the parameter of interest. Additionally, we show that these confidence intervals can also achieve second-order accuracy. We also show that the method is exact in three instances where the standard bootstrap fails. Using simulation studies, we illustrate the coverage accuracy of the method in three examples where standard parametric bootstrap procedures are computationally intensive and less accurate in finite samples.
\end{abstract}

\section{Introduction}
In many fields of research such as the social and life sciences, fitting parametric models to data and performing inference, in particular the construction of Confidence Intervals (CIs), is essential for decision-making. CIs are typically derived from the asymptotic distribution of a consistent estimator or through simulation-based methods such as the (parametric) bootstrap \citep{efron1979bootstrap}. However, these standard methods can be challenging to apply to complex parametric models due to the analytical and computational burden of obtaining consistent estimators. This particularly applies to models that incorporate specific features such as measurement disturbances (e.g., outliers, misclassification errors), complex dependence structures, missing data or censoring; see, e.g.,~\citet{zhang2023just} and~\citet{cavaliere2024bootstrap} for recent works with extensive examples.

Let $\vartheta_0\in\real$ denote the quantity of interest, related to the parameter $\bt_0\in\real^p$ of a model $F_{\bt_0}$ such that $\vartheta_0 \vcentcolon= \psi(\bt_0)$ for a known function $\psi$. Our primary objective is to construct valid CIs for $\vartheta_0$, derived from an estimator that is not required to be consistent. We have observations $Y_1, \dots, Y_n$ from $F_{\bt_0}$ and a simple, readily available but potentially inconsistent {\it initial estimator} $\hbpi_n\in\real^p$ based on these observations, i.e., $\hbpi_n$ converges in probability to $\bpi(\bt_0)\neq\bt_0$. In other words, the initial estimator has an asymptotic bias of $\bpi(\bt_0)-\bt_0$. Therefore, the estimator $\psi(\hbpi_n)$ is not necessarily consistent for $\vartheta_0$ and the standard bootstrap procedures on $\psi(\hbpi_n)$ do not provide valid CIs for $\vartheta_0$. A natural remedy is to correct $\hbpi_n$ so that it converges in probability to $\bt_0$ using, for instance, the framework of indirect inference~(see~\citealp{zhang2023just} and the references therein).  However, the proposed approach is designed so that such a correction is not necessary.

Given that generating simulated data is relatively straightforward in many parametric settings where inference is difficult, we aim to leverage this advantage and propose a novel simulation-based inference method called the {\it implicit bootstrap} that bypasses the need for a consistent estimator. This method is based on the  potentially inconsistent initial estimator $\hbpi_n$, which can be chosen for its analytical and numerical simplicity. The implicit bootstrap generates a distribution from which CIs for $\vartheta_0$ can be derived using the percentile method, which has the notable advantage of being transformation-respecting \citep{efron1994introduction}. While different, the implicit bootstrap shares similarities with generalized fiducial inference \citep{hannig2009generalized,hannig2016:GFIreview} and confidence distributions \citep[see, e.g.,][for a recent review]{BFF-bridge:2024}.

We summarize the key contributions of this paper as follows.
\begin{itemize}
    \item We demonstrate that the implicit bootstrap provides asymptotically valid CIs regardless of the potential inconsistency of the initial estimator. This result is derived under mild conditions, and allows, for example, the initial estimator to have a nonstandard limiting distribution. See~\cref{sec:1st:order}. 
    \item We demonstrate that, under similar requirements used to establish the second-order accuracy of the studentized bootstrap (see, e.g., \citealp{hall1992boot}), the implicit bootstrap provides second-order accurate percentile CIs. See~\cref{sec:2nd:order}. 
    \item We study cases in which the implicit bootstrap provides CIs with exact coverages for any sample sizes and nominal levels. See~\cref{sec:exact}.
\end{itemize}
As a byproduct, we derive a uniform version of Skorokhod's representation theorem that is deduced from a generalized Skorokhod's representation theorem proved in \cite{van1996weak}. We use this result to establish the asymptotic validity and second-order accuracy of the implicit bootstrap CIs. Additionally, we propose an alternative approach for computing the implicit bootstrap that has a computational cost comparable to the percentile parametric bootstrap. 

Consequently, the implicit bootstrap, which can be based on both consistent and inconsistent estimators, is a second-order accurate percentile method. In particular, when the initial estimator is consistent, the implicit bootstrap is an alternative to standard parametric bootstrap procedures. 

The rest of this article is organized as follows. In \cref{sec:methodology}, we formally introduce the implicit bootstrap. In \cref{sec:implicit-bootstrap}, we provide the theoretical justifications for its validity and we develop the sufficient conditions for establishing the second-order accuracy and the exact coverage property. In \cref{sec:simulation}, we present empirical evidence supporting the theoretical claims through simulation studies involving three parametric scenarios where inference is complex. The paper concludes in~\cref{sec:conclusion}.

\section{Methodology}\label{sec:methodology}
To introduce the implicit bootstrap in~\cref{sec:implicit_bootstrap}, we define the parametric setting and the initial estimator, specifying in what sense this estimator can be inconsistent. To provide intuition about the implicit bootstrap, we explain the insensitivity of the implicit bootstrap estimator to the inconsistency of the initial estimator in~\cref{sec:insensitivity}. 
\subsection{Implicit bootstrap}\label{sec:implicit_bootstrap}
Consider a random sample $\Y(\bt_0)\in\real^n$ generated from a parametric model $F_{\bt_0}$, where $\bt_0\in\real^p$. For the sake of simplicity, fixed and known quantities such as covariates are omitted from the notation. We recall that our goal is to construct valid CIs for $\vartheta_0\in\real$,  a differentiable transformation of $\bt_0$, defined as  $\vartheta_0\vcentcolon=\psi(\bt_0)$. We assume that the initial estimator $\hbpi_n$, computed on $\Y(\bt_0)$, converges in probability to a non-stochastic limit $\bpi(\bt_0)\neq\bt_0$. We suppose that there is a suitable subspace $\bT\subseteq\real^p$ such that $\bt_0\in\bT$.  We also assume that data can be simulated from $F_{\bt}$ for any $\bt\in\bT$. More specifically, we express the parametric data generating mechanism as $\Y(\bt) \vcentcolon= \mathbf{F}(\bt,\! \W)$ where $\W\in\bm{\mathcal{W}}\subset\real^m$ is a random vector whose distribution does not depend on $\bt\in\bT$. 
For example, if $Y_i(\bt), \, i = 1, \ldots, n$, are continuous independent and identically distributed (i.i.d.) random variables, we can write $\mathbf{F}(\bt,\! \W)\vcentcolon=\{F_\bt^{-1}(\text{W}_1),\dots,F_\bt^{-1}(\text{W}_m)\}^\top$ where $m = n$, $F_\bt^{-1}$ is the quantile function, $W_j \overset{i.i.d.}{\sim}U(0,1), \, j = 1, \ldots, n$, and $U(0,1)$ denotes the continuous uniform distribution on $(0,1)$. To avoid ambiguous definitions, we write $\hbpi_n=\hbpi_n(\bt_0,\W)$ and, for any $\bt\in\bT$, letting $\W^\ast$ correspond to an independent copy of $\W$, we denote $\hbpi_n(\bt,\W^\ast)$  the initial estimator based on the simulated random sample $\Y^\ast(\bt)\vcentcolon=\mathbf{F}(\bt,\! \W^\ast)$. We assume that, for any $\bt\in\bT$, $\hbpi_n(\bt,\W^\ast)$ converges in probability to $\bpi(\bt)$, where the function $\bpi(\cdot)$ is unknown and one-to-one. It is important to highlight that in practice, knowledge of $\W$ is not required, provided that simulated data $\Y(\bt)$ can be generated for any $\bt\in\bT$. Therefore, the introduction of $\W$ is primarily driven by the necessity for theoretical rigor, rather than being essential for the practical application of the method.

Let $\w,\w^\ast\in\bm{\mathcal{W}}$ be realizations of $\W$ and $\W^\ast$ respectively, we propose the following parametric approach. A realization of the {\it implicit bootstrap distribution} is given by solving the following matching equation
\begin{equation}
\widecheck\bt_n(\w^\ast|\w)\in\argmin_{\bt\in\bT} \; \|\hbpi_n(\bt_0,\w) - \hbpi_n(\bt,\w^\ast)\|,
    \label{eq:def:ImpBoot-estimator:bomegas}
\end{equation}
where $\|\cdot\|$ is the Euclidean norm. A direct consequence of the above definition is that if $\w=\w^\ast$, then $\bt_0$ is guaranteed to be a solution of the matching in \cref{eq:def:ImpBoot-estimator:bomegas}. For simplicity, we write $\hbpi_n^\ast(\bt)\vcentcolon=\hbpi_n(\bt,\W^\ast)$ and $\widecheck\bt_n^\ast\vcentcolon=\widecheck\bt_n(\W^\ast|\W)$, so that 
\begin{equation}\widecheck\bt_n^\ast\in \argmin_{\bt\in\bT} \; \|\hbpi_n - \hbpi_n^\ast(\bt)\|.
    \label{eq:def:ImpBoot-estimator}
\end{equation}

Therefore, the distribution of $\widecheck\bt_n^\ast$ is conditional on $\W$, or equivalently on $\hbpi_n$. We call $\widecheck\bt_n^\ast$ the {\it implicit bootstrap estimator} which we use to construct one-sided CIs for $\vartheta_0$ by taking, for any $\alpha\in (0,1)$,  the $\alpha$-level quantile of the distribution $\widecheck\vartheta^\ast_n\vcentcolon=\psi(\widecheck\bt_n^\ast)$, i.e., 
\begin{equation*}
\widecheck{\vartheta}^\ast_{n,\alpha}\vcentcolon=\inf\left\{x\in\real \mid \Pr\left(\widecheck\vartheta_n^\ast\leq x \mid \hbpi_n\right)\geq\alpha\right\}.
\end{equation*}
For the rest of the paper, we refer to the implicit bootstrap when using the percentile method on the distribution of $\widecheck\vartheta_n^\ast$ to construct CIs for $\vartheta_0$.

The implementation of the implicit bootstrap is similar to parametric bootstrap procedures. The computation of $\widecheck\bt^\ast_n$ can be directly obtained from its definition in~\cref{eq:def:ImpBoot-estimator} for a large number of simulated samples. To improve computational efficiency, we propose an alternative approach for initial estimators $\hbpi_n$ that can be expressed as Z-estimators. This approach has a computational cost comparable to the percentile parametric bootstrap based on $\hbpi_n$.

\subsection{Insensitivity to estimator inconsistency}\label{sec:insensitivity}
One of the main assets of the implicit bootstrap is its insensitivity to the inconsistency of the initial estimator, that is, the validity of the method is not tied to the consistency of $\hbpi_n$. As an illustration, in the case where the function $\bpi(\bt)$ is a bijection with continuous inverse and \cref{eq:def:ImpBoot-estimator} corresponds to an exact solution, i.e., $\widecheck\bt_n^\ast\in \argzero_{\bt\in\bT} \; \|\hbpi_n - \hbpi_n^\ast(\bt)\|$, we have
\begin{equation*}
    \widecheck\bt_n^\ast\in \argzero_{\bt\in\bT} \; \|\hbpi_n - \hbpi_n^\ast(\bt)\| = \argzero_{\bt\in\bT} \; \|\bpi^{-1}(\hbpi_n) - \bpi^{-1}\{\hbpi_n^\ast(\bt)\}\|.
\end{equation*}
In other words, in this specific instance, we can assume without loss of generality that the implicit bootstrap procedure is based on the consistent estimator $\bpi^{-1}(\hbpi_n)$ while only requiring to compute the inconsistent $\hbpi_n$. In the general case, the insensitivity of the implicit bootstrap to the inconsistency of the initial estimator is related to the existence of the following minimum distance estimator (or generalized method of moments estimator) defined as 
\begin{equation}\label{def:ideal:estim:general}
\widehat\bt_n\in\argmin_{\bt\in\bT}\|\hbpi_n-\bpi(\bt)\|. 
\end{equation}
This corrected estimator can only be considered theoretically, as the function $\bpi(\bt)$ is unknown. For this reason, we call $\widehat\bt_n$ the {\it idealized estimator}. Even though the implicit bootstrap does not require the knowledge of $\widehat\bt_n$, the introduction of this estimator allows to develop theoretical understanding for the validity of our method. As an illustration, under conditions assumed for the rest of this section, we can write $\widehat\bt_n\vcentcolon=\bpi^{-1}(\hbpi_n)$. If $\bpi^{-1}(\cdot)$ is sufficiently smooth, then we obtain the consistency of $\widehat\bt_n$ for $\bt_0$ from the continuous mapping theorem. Moreover, if the initial estimator is asymptotically normally distributed, so will be the idealized estimator, by the delta method. Namely, if  
\begin{equation*}
n^{\nicefrac{1}{2}}\bm\Sigma(\bt_0)^{-\nicefrac{1}{2}}\{\hbpi_n - \bpi(\bt_0)\}\dovr\mathcal{N}(\bm{0},\I_p), 
\end{equation*}
then
\begin{equation*}
   n^{\nicefrac{1}{2}}\bSig(\bt_0)^{-\nicefrac{1}{2}}\bm{A}(\bt_0)\left(\widehat\bt_n - \bt_0\right)\dovr \mathcal{N}(\bm{0},\I_p), 
\end{equation*}
where $\bm{0}$ is the zero vector in $\real^p$, $\I_p$ is the $p \times p$ identity matrix, $\bSig(\bt_0)$ is the covariance matrix of $n^{\nicefrac{1}{2}}\hbpi_n$ and $\bm{A}(\bt_0)$ is the Jacobian matrix of $\bpi(\bt)$ at $\bt_0$. Let $\widehat\bt_n^\ast\vcentcolon=\bpi^{-1}\{\hbpi_n^\ast(\widehat\bt_n)\}$ be the bootstrapped idealized estimator and $\bm\Xi(\bt)\vcentcolon=\bSig(\bt)^{-\nicefrac{1}{2}}\bm{A}(\bt)$, then it can be shown that
\begin{equation*}
     n^{\nicefrac{1}{2}} \bm\Xi(\widehat{\bt}_n)  \left( \widehat{\bt}_n^\ast  - \widehat{\bt}_n\right) \overset{d}{\to} \mathcal{N}(\bm{0},\I_p),
\end{equation*}
in probability. Under the same conditions, we can also show that
\begin{equation*}
    n^{\nicefrac{1}{2}} \bm\Xi(\widehat{\bt}_n) \left(\widecheck\bt_n^\ast  - \widehat\bt_n\right) \overset{d}{\to} \mathcal{N}(\bm{0},\I_p) 
\end{equation*}
in probability. Therefore, the distribution of $\widecheck\bt_n^\ast$ is asymptotically equivalent to the distribution of the bootstrap estimator $\widehat{\bt}_n^\ast$ of the idealized estimator $\widehat{\bt}_n$. This property, as well as the fact that $\widehat{\bt}_n^\ast$ can only be considered implicitly or theoretically, explain and justify the chosen name for the estimator $\widecheck\bt_n^\ast$ and the method.

\section{Theoretical properties}\label{sec:implicit-bootstrap}
In this section, we study the coverage properties of the implicit bootstrap, i.e., we give sufficient conditions to characterize the error term $e$ in $\Pr(\vartheta_0\leq\widecheck{\vartheta}^\ast_{n,\alpha})=\alpha + e$. First, we develop the conditions to demonstrate that our approach is valid, i.e., $e=o(1)$, and second-order accurate, i.e., $e=o(n^{-\nicefrac{1}{2}})$. Naturally, the assumptions required for the second-order accuracy of the implicit bootstrap imply those imposed for its validity. Finally, we develop a non-asymptotic theory to prove that the implicit bootstrap can achieve exact coverage, i.e., $e=0$. All these results carry over to two-sided CIs. Throughout this section, we will refer to the idealized estimator $\widehat\bt_n$, introduced in \cref{sec:insensitivity}, to discuss the required assumptions for our approach given the relationship between the implicit bootstrap estimator $\widehat\bt^\ast_n$ and the bootstrap estimator $\widehat\bt^\ast_n$. As previously mentioned, $\widehat\bt_n$ is not available in our framework since $\bpi(\bt)$ is unknown and it is unnecessary to implement the implicit bootstrap.
\subsection{Validity}\label{sec:1st:order} 
In this section, we study the validity of the implicit bootstrap. We begin by introducing our first assumption, which specifies the conditions on $\bT$.
\setcounter{Assumption}{0}
\renewcommand\theAssumption{A.\arabic{Assumption}}
\begin{Assumption}\label{assum:Theta_1:convex:compact}
The space $\bT$ is compact, convex and such that $\bt_0\in\Int(\bT)$.
\end{Assumption}
\cref{assum:Theta_1:convex:compact} is a common regularity condition often used to establish the consistency and asymptotic normality of extremum  estimators (see, e.g.,  \citealp{white1982maximum, newey1994large, van1998asymptotic}). In specific instances, the space $\bT$ may either correspond to the whole parameter space or a (small) neighborhood of $\bt_0$. The primary purpose of this assumption is to ensure that expansions can be made among interior points of $\bT$ and to guarantee the boundedness of continuous functions defined on $\bT$. When $\hbpi_n(\bt)$ has a closed-form expression, as discussed in \cref{sec:exact}, less restrictive conditions may suffice. However, we do not attempt to pursue the weakest possible conditions to avoid overly technical treatments in establishing the theoretical results.

The next assumption imposes smoothness requirements on the functions $\psi(\bt)$ and $\bpi(\bt)$. 
\begin{Assumption}\label{assum:first:order:asymp_bias:function}
The functions $\psi(\bt)$ and $\bpi(\bt)$ are continuously differentiable. Moreover, $\bpi(\bt)$  is one-to-one on $\bT$ with Jacobian matrix $\bm{A}(\bt)\vcentcolon=\partial\bpi(\bt)/\partial\bt^\top$ having full rank for any $\bt$ in an open set containing $\bT$.
\end{Assumption}

\cref{assum:first:order:asymp_bias:function} is plausible for a wide range of models and estimators. It implies, in particular, that $\bt_0$ is identifiable since $\bpi(\bt)$ is required to be one-to-one on $\bT$. Primitive conditions that guarantee that $\bpi(\bt)$ is one-to-one can be found, for example, in \cite{komunjer2012global} and references therein. Furthermore, under \cref{assum:Theta_1:convex:compact} and since $\bpi(\bt)$ is continuous on $\bT$ under \cref{assum:first:order:asymp_bias:function}, the idealized estimator defined in \cref{def:ideal:estim:general} is uniquely defined, that is,
\begin{equation}\label{def:ideal:estim}
\widehat\bt_n\vcentcolon=\argmin_{\bt\in\bT}\|\hbpi_n-\bpi(\bt)\|. 
\end{equation}
Similarly, for any $\bt\in\bT$, we can define $ \widehat\bt_n(\bt)\vcentcolon=\argmin_{\bt_1\in\bT}\|\hbpi_n(\bt)-\bpi(\bt_1)\|$. If the support of $\hbpi_n$ and $\hbpi_n(\bt)$ are contained in $\bpi(\bT)$, we can simply write $\widehat\bt_n=\bpi^{-1}(\hbpi_n)$ and $\widehat\bt_n(\bt)=\bpi^{-1}\{\hbpi_n(\bt)\}$. Using Theorem 3.2 in \cite{newey1994large}, the differentiability requirements in  \cref{assum:first:order:asymp_bias:function} are among the conditions ensuring the asymptotic normality of $\widehat\bt_n$ and $\widehat\bt_n(\bt)$. Regarding the implicit bootstrap estimator, these requirements are among conditions ensuring its weak convergence. These conditions are satisfied, for example, when the data exhibit features such as censoring, with the initial estimator being the ordinary least square estimator (see, e.g., \citealp{greene1981asymptotic}). These conditions are also often considered, for example, in indirect inference (see, e.g., \citealp{smith1993estimating, gourieroux1993indirect, gallant1996moments}). Moreover, if $\bpi(\bt)$ is assumed to be additive, i.e., $\bpi(\bt)=\mathbf{B}+\bt$, as in~\citet{cavaliere2024bootstrap},  then~\cref{assum:first:order:asymp_bias:function} is satisfied. Finally, it is trivially satisfied if the estimator is consistent since, in that case, $\bpi(\bt)=\bt$. 

The third assumption describes the limiting distribution of the initial estimator.

\begin{Assumption}\label{assum:first:order:highlevel}
There exists a diverging sequence of positive numbers $\{a_n\}_{n\geqslant 1}$ such that we have, uniformly in $\bt\in\bT$, 
\begin{equation*}
a_n\bm\Sigma_n(\bt)^{-\nicefrac{1}{2}}\{\hbpi_n(\bt) - \bpi(\bt)\}\dovr\mathbf{Z},
\end{equation*}
where $\bm\Sigma_n(\bt)\vcentcolon=\cov\{a_n\hbpi_n(\bt)\}$ is the covariance and $\mathbf{Z}$ is a zero-mean random vector whose distribution is continuous and does not depend on $\bt\in\bT$. Moreover, the sequence $\{\bm\Sigma_n(\bt)\}_{n\geq 1}$ converges uniformly in $\bt\in\bT$ to a positive definite matrix $\bm\Sigma(\bt)$ which is continuous in $\bt\in\bT$.
\end{Assumption}
\cref{assum:first:order:highlevel} requires, in particular, that the limiting distribution of $\hbpi_n(\bt)$ is determined for any $\bt\in\bT$. This is a common regularity condition ensuring, for example, that the limiting distribution of the idealized estimator $\widehat\bt_n(\bt)$ is determined for any $\bt\in\bT$ (see, e.g., Theorem 3.2 in \citealp{newey1994large}). In the particular case where the support of $\hbpi_n(\bt)$ is contained in $\bpi(\bT)$ for any $\bt\in\bT$ and using Assumptions~\ref{assum:first:order:asymp_bias:function} and \ref{assum:first:order:highlevel}, the limiting distributions of $\widehat\bt_n(\bt)$ is simply determined using the delta method.

The main requirement in \cref{assum:first:order:highlevel} is that the convergence in distribution of $\hbpi_n(\bt)$ holds uniformly in $\bt\in\bT$. Generally, uniform convergence requirements are regularity conditions imposed to demonstrate the validity of parametric statistical procedures (see, e.g., \citealp{newey1991uniform,newey1994large,phillips2012folklore,arvanitis2018validity, kasy2019uniformity}). Although not necessarily the weakest, \cref{assum:first:order:highlevel} can be viewed as a primitive condition ensuring that standard parametric bootstrap methods to construct CIs are valid. Indeed, the uniformity requirement can be equivalently expressed as follows: for any sequence $\{\x_n\}_{n\geq1}\subset\bT$, 
\begin{equation}\label{eq:unif:conv:distrib:equiv:def}
a_n\bm\Sigma_n(\x_n)^{-\nicefrac{1}{2}}\{\hbpi_n(\x_n) - \bpi(\x_n)\}\dovr\mathbf{Z}.
\end{equation}
If the initial estimator $\hbpi_n(\bt)$ is consistent for $\bt\in\bT$, i.e., $\bpi(\bt)=\bt$, then \cref{eq:unif:conv:distrib:equiv:def} directly implies that the requirements of Lemma 23.3 in~\citealp{van1998asymptotic} hold. This result ensures the validity of the studentized bootstrap based on this (consistent) estimator $\hbpi_n(\bt)$ and, provided that $\Z$ is symmetrically distributed around zero, the validity of the percentile bootstrap based on $\hbpi_n(\bt)$. The validity of the implicit bootstrap does not rely on the consistency of the initial estimator nor on the symmetry of the limiting distribution $\Z$ (see \cref{assum:first:order:highlevel}).

\begin{Theorem}\label{thm:first:order}
Let $\alpha\in(0,1)$. Then, under Assumptions~\ref{assum:Theta_1:convex:compact},~\ref{assum:first:order:asymp_bias:function} and~\ref{assum:first:order:highlevel},~we~have, 
\begin{equation*}
    \Pr\left(\vartheta_0\leq\widecheck{\vartheta}^\ast_{n,\alpha}\right)=\alpha+o(1).
\end{equation*}
\end{Theorem}
\cref{thm:first:order} ensures the validity of the implicit bootstrap under common regularity conditions. Even though the idealized estimator can only be theoretically considered in our framework, the studentized bootstrap based on $\widehat\vartheta_n\vcentcolon=\psi(\widehat\bt_n)$ can be demonstrated to provide valid CIs for $\vartheta_0$ under the assumptions of \cref{thm:first:order}. Provided that $\Z$ is symmetrically distributed around zero, the same conclusion holds for the percentile bootstrap based on $\widehat\vartheta_n$. Our method, on the other hand, does not require the symmetry of $\Z$. 

One interesting result that we derive to prove~\cref{thm:first:order} is a uniform extension of Skorokhod's representation theorem that applies to $\X_n(\bt)\vcentcolon=a_n\bSig_n(\bt)^{-\nicefrac{1}{2}}\{\hbpi_n(\bt) - \bpi(\bt)\}$ and $\Z$.

\subsection{Second-order accuracy}\label{sec:2nd:order}

\setcounter{Assumption}{0}
\renewcommand\theAssumption{B.\arabic{Assumption}}

In this section, we study the second-order accuracy of the implicit bootstrap. We begin with Assumptions~\ref{assum:2nd:order:asymp_bias:function} and~\ref{assum:2nd:order:highlevel}, which are stronger than Assumptions~\ref{assum:first:order:asymp_bias:function} and \ref{assum:first:order:highlevel}, respectively. These assumptions ensure that the error term 
\begin{equation}\label{eq:def:matching:error:term}
    \delta_n\vcentcolon=\|\hbpi_n - \hbpi_n^\ast(\widecheck\bt_n^\ast)\|,
\end{equation}
is such that $\delta_n=o_p(n^{-1})$. Its role for establishing the second-order accuracy of the implicit bootstrap is discussed in \cref{rem:second-order:correct:critical:point}. Nevertheless, it is important to note that if a perfect matching is available, i.e.,  if $\widecheck\bt_n^\ast$ is such that $\delta_n=0$, then Assumptions~\ref{assum:2nd:order:asymp_bias:function} and~\ref{assum:2nd:order:highlevel} below are not necessary. 
\begin{Assumption}\label{assum:2nd:order:asymp_bias:function}%
The functions $\psi(\bt)$ and $\bpi(\bt)$ are twice continuously differentiable. Moreover, $\bpi(\bt)$, restricted to $\bT$, has a continuously differentiable inverse and the Jacobian matrix $\bm{A}(\bt)\vcentcolon=\partial\bpi(\bt)/\partial\bt^\top$ has full rank for any $\bt$ in an open set containing $\bT$. 
\end{Assumption}

We immediately see that \cref{assum:2nd:order:asymp_bias:function} implies \cref{assum:first:order:asymp_bias:function}. The essential distinction between these two assumptions is that the former allows us to make use of the mean value theorem with the map $\bpi^{-1}(\cdot)$. Primitive conditions ensuring that $\bpi(\bt)$ has a continuous inverse can be found in \citealp{komunjer2012global} and references therein, one of which being that $\bpi(\bt)$ is twice continuously differentiable.

Supposing that \cref{assum:first:order:highlevel} holds with $a_n=n^{\nicefrac{1}{2}}$, a uniform extension of Skorokhod's representation theorem applies to $\X_n(\bt)=n^{\nicefrac{1}{2}}\bSig_n(\bt)^{-\nicefrac{1}{2}}\{\hbpi_n(\bt) - \bpi(\bt)\}$ and $\Z$. In particular, we can assume, without loss of generality, that a common sample space $\bOmega$ can be considered for $\X_n(\bt)$ and $\Z$ and that $\X_n(\bt)$ converges almost surely to $\Z$ uniformly in $\bt\in\bT$. Consequently, for almost all $\bomega\in\bOmega$, we have $\|\X_n(\bt,\bomega)-\Z(\bomega)\| = o(1)$ uniformly in $\bt\in\bT$. In the following assumption, we consider $\mathcal{R}_n(\bt,\bomega)\vcentcolon=\X_n(\bt,\bomega)-\Z(\bomega)$ and require a specific order for $\|\mathcal{R}_n(\bt, \bomega) \|$ and a continuity property.
\begin{Assumption}\label{assum:2nd:order:highlevel}
Suppose~\cref{assum:first:order:highlevel} holds and the components of the matrix $\bSig(\bt)$ are continuously differentiable in $\bt\in\bT$. Moreover,  $\|\mathcal{R}_n(\bt, \bomega) \|=~\mathcal{O}(n^{-\nicefrac{1}{2}})$, for~almost all $\bomega\in\bOmega$, uniformly in $\bt\in\bT$ and the family $\{n^{\nicefrac{1}{2}}\mathcal{R}_n(\bt,\bomega)\}_{n\geq 1}$ is equicontinuous in $\bt\in\bT$.
\end{Assumption}

Since $\bT$ is compact by \cref{assum:Theta_1:convex:compact}, the equicontinuity requirement on $\{n^{\nicefrac{1}{2}}\mathcal{R}_n(\bt,\bomega)\}_{n\geq1}$ implies that this family is uniformly equicontinuous on $\bT$, for almost all $\bomega\in\bOmega$. Therefore, for any sequences $\{\x_n\}_{n\geq1},\{\y_n\}_{n\geq1}\subset\bT$ such that $\|\x_n-\y_n\|=o(1)$, we have $\|\mathcal{R}_n(\x_n,\bomega)- \mathcal{R}_n(\y_n,\bomega)\|=o(n^{-\nicefrac{1}{2}})$, for almost all $\bomega\in\bOmega$. Similarly to uniformity, (stochastic) equicontinuity requirements are common regularity conditions used in parametric statistical procedures (see, e.g., \citealp{newey1991uniform} or \citealp{andrews1992generic}).

The last two assumptions of this section involve $\widehat\vartheta_n=\psi(\widehat\bt_n)$ as well as a bootstrap version of it. More precisely, Assumptions \ref{assum:2nd:order:EE:high:level} and \ref{assum:2nd:order:Hall:delta:method} below, imply that the studentized bootstrap based on $\widehat\vartheta_n$ is second-order accurate. These assumptions play similar roles for the implicit bootstrap, although both $\widehat\vartheta_n$ and $\widehat\bt_n$ can only be considered theoretically. 
Before stating Assumptions \ref{assum:2nd:order:EE:high:level} and \ref{assum:2nd:order:Hall:delta:method}, we need to define some standardized random variables.
Supposing that \cref{assum:first:order:highlevel} holds with $a_n=n^{\nicefrac{1}{2}}$ and $\mathbf{Z}$ is a random vector following a standard multivariate normal distribution, we can deduce that, conditionally on $\hbpi_n$,
\begin{equation}\label{def:stand:stat:imp:boot}
 \widecheck\xi_n^\ast\vcentcolon=  n^{\nicefrac{1}{2}}(\widecheck\vartheta^\ast_n - \widehat\vartheta_n)/\widecheck\sigma^\ast \dovr \mathcal{N}(0,1),
\end{equation}
in probability, where $\widecheck\sigma^\ast$ is a consistent estimator of the square root of the variance of $n^{\nicefrac{1}{2}}\widecheck\vartheta^\ast_n$. \cref{assum:2nd:order:EE:high:level} involves the following studentized statistics that are asymptotically normally distributed under the same conditions,
\begin{align}\label{def:student:stat:&:boot:student:stat}
\begin{split}
    \widehat\zeta_n&\vcentcolon=n^{\nicefrac{1}{2}}(\widehat\vartheta_n-\vartheta_0)/\widehat\sigma, \\
    \widehat\zeta_n^{\ast\ast}&\vcentcolon=n^{\nicefrac{1}{2}}(\widehat\vartheta_n^{\ast\ast} - \widecheck\vartheta^\ast_n)/\widehat\sigma^{\ast\ast},
\end{split}
\end{align}
where  $\widehat\vartheta_n^{\ast\ast}\vcentcolon=\psi(\widehat\bt_n^{\ast\ast})$, $\widehat\bt_n^{\ast\ast}\vcentcolon=\argmin_{\bt\in\bT}\|\hbpi_n^{\ast\ast}(\widecheck\bt_n^{\ast}) - \bpi(\bt)\|$, $\hbpi_n^{\ast\ast}(\widecheck\bt_n^{\ast})$ is based on the simulated sample $\Y^{\ast\ast}(\widecheck\bt_n^{\ast}) = \mathbf{F}(\widecheck\bt_n^{\ast}, \W^{\ast\ast})$ with $\W^{\ast\ast}$ being an independent copy of~$\W$ and $\widehat\sigma$ and $\widehat\sigma^{\ast\ast}$ are consistent estimators of the square root of the variances of $n^{\nicefrac{1}{2}}\widehat\vartheta_n$ and $n^{\nicefrac{1}{2}}\widehat\vartheta_n^{\ast\ast}$, respectively. 
In what follows, we denote by $\widecheck\xi_{n,\alpha}^\ast, \widehat\zeta_{n,\alpha}$ and $\widehat\zeta_{n,\alpha}^{\ast\ast}$ the $\alpha$-level quantile of $\widecheck\xi_n^\ast, \widehat\zeta_n$ and $\widehat\zeta_n^{\ast\ast}$, respectively.

\begin{Assumption}\label{assum:2nd:order:EE:high:level}
We have uniformly in $x\in\real$,
\begin{align}\label{eq:EE:vartheta}
\begin{split}
\Pr\left(\widehat\zeta_n \leq x \right) &= \Phi(x) + n^{-\nicefrac{1}{2}}q(x)\phi(x) + o\left(n^{-\nicefrac{1}{2}}\right),
\\
\Pr\left(\widehat\zeta_n^{\ast\ast} \leq x  \big| \widecheck\bt^{\ast}_n\right) &= \Phi(x) + n^{-\nicefrac{1}{2}}\widehat{q}(x)\phi(x) + o_p\left(n^{-\nicefrac{1}{2}}\right),
\end{split}
\end{align}
where $\Phi,\phi$ are the distribution and density functions of a standard normal random variable, $q(x)$ is an even polynomial in~$x$, $\widehat{q}(x) = q(x) + o_p(1)$, and for some constant $c>\nicefrac{1}{2}$, 
\begin{align*}
    \widehat\zeta_{n,\alpha}&=z_\alpha - n^{-\nicefrac{1}{2}}q(z_\alpha)+o(n^{-\nicefrac{1}{2}}), \\
    \widehat\zeta_{n,\alpha}^{\ast\ast}&=z_\alpha - n^{-\nicefrac{1}{2}}\widehat{q}(z_\alpha)+o_p(n^{-\nicefrac{1}{2}}), 
\end{align*}
uniformly in $\alpha\in[n^{-c},1-n^{-c}]$, where $z_\alpha\vcentcolon=\Phi^{-1}(\alpha)$.
\end{Assumption}
\cref{assum:2nd:order:EE:high:level} states that the studentized statistics of $\widehat{\vartheta}_n,\widehat{\vartheta}_n^{\ast\ast}$ have an Edgeworth and Cornish-Fisher expansions up to an order $n^{-\nicefrac{1}{2}}$. Cornish-Fisher expansions often hold under the same conditions as Edgeworth expansions. The estimator $\widehat\vartheta_n$ is a smooth transformation of the initial estimator $\hbpi_n$ when the support of the latter is contained in $\bpi(\bT)$, since in that case we can write $\widehat\vartheta_n = \psi\{\bpi^{-1}(\hbpi_n)\}$, by \cref{assum:2nd:order:asymp_bias:function}. The validity of Edgeworth expansion under smooth transformation is discussed in~\citet{bhattacharya1978validity} and \citet{hall1992boot}. The key requirements are a valid Edgeworth expansion for $\hbpi_n$ and some smoothness conditions on the transformation (\citealp{skovgaard1981transformation}). The exact requirements for an Edgeworth expansion for $\hbpi_n$ depend on the estimator's form. Extensive literature covers the validity of  Edgeworth expansion for various statistics, such as averages of independent random vectors~\citep{petrov1975sums,bhattacharya1976normal}, smooth function of averages of transformations of random vectors~\citep{bhattacharya1978validity,skovgaard1981transformation,hall1992boot}, and the Maximum Likelihood Estimator (MLE)~\citep{chibisov1974asymptotic,bhattacharya1978validity,skovgaard1981edgeworth,lieberman2003valid}; other studies on the topic include~\citep{chambers1967methods,sargan1976econometric,phillips1977general,lahiri2010edgeworth,arvanitis2018validity}. 

To demonstrate that studentized bootstrap CIs are second-order accurate, an Edgeworth expansion for the bootstrapped studentized statistics of $\widehat{\vartheta}_n$, i.e., $n^{-\nicefrac{1}{2}}(\widehat\vartheta_n^\ast-\widehat\vartheta_n)/\widehat\sigma^\ast$, conditionally on $\widehat\bt_n$, is typically necessary. The condition on $\widehat\vartheta^{\ast\ast}_n$ in~\cref{assum:2nd:order:EE:high:level} mirrors this requirement and can be viewed as a second parametric bootstrap iteration, where $\widehat\bt_n^\ast$ is replaced by $\widecheck\bt_n^\ast$. Informally, this hypothesis holds by the \textit{bootstrap principle} (\citealp{hall1992boot}). In the parametric context, it can be rigorously established if a valid Edgeworth expansion for the studentized statistics of $\widehat\vartheta_n(\bt)\vcentcolon=\psi\{\widehat\bt_n(\bt)\}$  holds uniformly in $\bt\in\bT$ (see, e.g., Assumption~5 in~\citealp{arvanitis2018validity} and our discussion on~\cref{assum:first:order:highlevel}).

The parity of the (population) polynomial $q$ as well as the convergence of $\widehat{q}$ to  $q$ are common assumptions in the theory of second-order accurate coverage of the studentized bootstrap (see \citealp{hall1992boot}). 
In the parametric context, $q$ may be viewed as a function of $\bt$, and the convergence of $\widehat{q}$ to $q$ can be justified by the consistency of $\widecheck\bt^\ast_n$ and the continuous mapping theorem.

\setcounter{Assumption}{3}
\renewcommand\theAssumption{B.\arabic{Assumption}}

As discussed in \cref{rem:second-order:correct:critical:point}, Assumptions~\ref{assum:Theta_1:convex:compact}, \ref{assum:2nd:order:asymp_bias:function}, \ref{assum:2nd:order:highlevel} and \ref{assum:2nd:order:EE:high:level} allow us to establish the  following expansion:  
\begin{equation}\label{eq:2nd:check:vartheta:quantile:expansion}
    \widecheck\xi_{n,\alpha}^\ast = z_\alpha + n^{-\nicefrac{1}{2}}q(z_\alpha)+\Delta,
\end{equation}
where $\Delta=o_p(n^{-\nicefrac{1}{2}})$ and $\alpha\in(0,1)$. The next assumption imposes requirements on the remainder term $\Delta$.

\begin{Assumption}\label{assum:2nd:order:Hall:delta:method}
Let $\Delta$ be the remainder term in \cref{eq:2nd:check:vartheta:quantile:expansion}. Then, for any $x\in\real$, we have
    \begin{equation*}
        \Pr\left(\widehat\zeta_n + \Delta \leq x\right) = \Pr\left(\widehat\zeta_n  \leq x\right) + o(n^{-\nicefrac{1}{2}}).
    \end{equation*}
\end{Assumption}
\cref{assum:2nd:order:Hall:delta:method} essentially states that $\widehat\zeta_n+\Delta$ has an Edgeworth expansion that is $o(n^{-\nicefrac{1}{2}})$ close to the Edgeworth expansion of $\widehat\zeta_n$.  Sufficient conditions can be found in \citealp[Section 2.7]{hall1992boot}. This condition is typically required for the studentized bootstrap CIs to be second-order accurate (see, e.g., Proposition 3.1 and Section 3.4.5 in \citealp{hall1992boot}).
\begin{Theorem}\label{thm:second:order}
Let $\alpha\in(0,1)$. Then, under Assumptions~\ref{assum:Theta_1:convex:compact}, \ref{assum:2nd:order:asymp_bias:function}, \ref{assum:2nd:order:highlevel}, \ref{assum:2nd:order:EE:high:level} and \ref{assum:2nd:order:Hall:delta:method}, we have,
\begin{equation*}
\Pr\left(\vartheta_0\leq\widecheck{\vartheta}^\ast_{n, \alpha}\right)=\alpha+o(n^{-\nicefrac{1}{2}}).
\end{equation*}  
\end{Theorem}
The key finding of this result is that when the error term in~\cref{eq:def:matching:error:term} is small enough, i.e., $\delta_n=o_p(n^{-1})$, the implicit bootstrap is second-order accurate under similar assumptions than required for the studentized bootstrap based on $\widehat\vartheta_n$ to be second-order accurate. Moreover, \cref{thm:second:order} ensures that the implicit bootstrap is a percentile approach with second-order accurate CIs that does not  necessitate any form of adjustments on $\widecheck\vartheta_{n,\alpha}^\ast$. This feature is particularly appealing for a general-purpose bootstrap method \citep{diciccio1995bootstrap}.

\begin{Remark}\label{rem:second-order:correct:critical:point}
We show that, under Assumptions~\ref{assum:Theta_1:convex:compact}, \ref{assum:2nd:order:asymp_bias:function}, \ref{assum:2nd:order:highlevel} and \ref{assum:2nd:order:EE:high:level}, we have the following asymptotic expansion 
\begin{equation*}
     \Pr\left(\widecheck\xi_n^\ast \leq x \big| \hbpi_n\right) = \Phi(x) - n^{-\nicefrac{1}{2}}q(x)\phi(x) + o_p\left(n^{-\nicefrac{1}{2}}\right).
\end{equation*}  
As a consequence, if $\delta_n=o_p(n^{-1})$, we obtain, for $\alpha \in(0,1)$, an expansion for $\widecheck\xi_{n,\alpha}^\ast$ that is, as \citet{hall1992boot} describes, ``second-order correct relative to'' the $\alpha$-level quantile of $-\widehat\zeta_n$. Second-order correctness of $\widecheck\xi_{n,\alpha}^\ast$ is fundamental for demonstrating the second-order accuracy of the implicit bootstrap in~\cref{thm:second:order}.  Moreover, as previously mentioned, if $\delta_n=0$, Assumptions \ref{assum:2nd:order:asymp_bias:function} and \ref{assum:2nd:order:highlevel} can be removed from the requirements of~\cref{thm:second:order}. 
\end{Remark}

\subsection{Exact coverage}\label{sec:exact}
\setcounter{Assumption}{0}
\renewcommand\theAssumption{C}
In this section, we show that the implicit bootstrap achieves exact coverage under a specific form for $\widecheck{\vartheta}^\ast_n$. Recall that, by definition of the initial estimator $\hbpi_n(\bt)$ and the implicit bootstrap estimator in~\cref{eq:def:ImpBoot-estimator:bomegas}, there exists a random vector $\W$, with support $\bm{\mathcal{W}}\subset\real^m$  such that we can write $\hbpi_n(\bt) = \hbpi_n(\bt, \W)$ and $\widecheck\bt_n^\ast = \widecheck\bt_n(\W^\ast|\W)$, where $\W^\ast$ is an independent copy of $\W$. The next assumption imposes some requirements on $\W$.

\begin{Assumption}\label{assum:exact:psi:factor}
The random vector $\W$ is absolutely continuous. There exists a continuous function $h:\bcalW\to\real$, and a set of univariate functions $\{h_{\w}\}_{\w\in\bcalW}$ continuous and strictly increasing, for almost all $\w\in\bcalW$ (or strictly decreasing for almost all $\w\in\bcalW$) such that 
    \begin{equation*}
            \widecheck\vartheta_n(\w^\ast|\w) 
        \vcentcolon=  \psi\{\widecheck\bt_n(\w^\ast|\w)\} 
        =  h_{\w}\{ h(\w^\ast)\},
    \end{equation*}
and $\widecheck\vartheta_n(\w|\w)=\vartheta_0$ for almost all $\w,\w^\ast\in\bcalW$.
\end{Assumption}
A direct consequence of \cref{assum:exact:psi:factor} is that $\widecheck\vartheta_n^\ast$ is an absolutely continuous random variable. \cref{assum:exact:psi:factor} may be verified when $\widecheck\vartheta_n^\ast$ is known in closed-form. A primitive condition for this assumption to hold is the applicability of a global implicit function theorem to \cref{eq:def:ImpBoot-estimator:bomegas}, for almost all $\w\in\bcalW$ (see, e.g., ~\citealp{sandberg1981global},~\citealp{idczak2016generalization} or~\citealp{cristea2017global} for sufficient conditions). Hence, the requirements imposed for achieving exact coverage with the implicit bootstrap may significantly differ from those imposed for the validity and the second-order accuracy of the method. In particular, the conditions that $\bT$ is compact and $\bt_0$ is in the interior of $\bT$ (see \cref{assum:Theta_1:convex:compact}) may be relaxed, as illustrated in \cref{eg:unif} to \cref{Example:Andrews}.

\cref{thm:exact} states that the implicit bootstrap has exact coverage.

\begin{Theorem}\label{thm:exact}
Let $\alpha\in(0,1)$, then under 
\cref{assum:exact:psi:factor}, we have $\Pr\left(\vartheta_0\leq\widecheck{\vartheta}^\ast_{n, \alpha}\right)=\alpha$.
\end{Theorem}

In what follows, we consider three examples that use the results developed for the exact coverage. In the first and second examples, we demonstrate that \cref{assum:exact:psi:factor} holds, hence \cref{thm:exact} applies. \cref{eg:unif} is a well-known example discussed in \citet{bickel1981some} for which, in particular, the percentile parametric bootstrap is not valid. \cref{example:Pareto} illustrates exact coverage for the parameters of the Pareto distribution. Although, by \cref{assum:exact:psi:factor}, $\widecheck\vartheta_n^\ast$ is necessarily an absolutely continuous random variable, the implicit bootstrap can still achieve exact coverage in a broader sense when $\widecheck\vartheta_n^\ast$ is not absolutely continuous. This is illustrated in \cref{Example:Andrews} where we show that $\Pr(\vartheta_0\leq\widecheck{\vartheta}^\ast_{n, \alpha})\geq\alpha$ for any $\alpha\in(0,1)$. 
Interestingly, \cref{Example:Andrews} is a case presented in \citet{Andrews:00} for which neither the parametric nor the nonparametric bootstrap is valid.

\begin{Example}[Upper bound of the uniform distribution]\label{eg:unif}
    Suppose that $Y_1,\dots,Y_n$ are i.i.d. $U(0,\theta_0)$, $\theta_0>0$, and the initial estimator is the sample maximum: $\widehat{\pi}_n=Y_{(n)}$, where $Y_{(1)}\leq\dots\leq Y_{(n)}$ denotes the order statistics of the sample $Y_1,\dots,Y_n$. Here, the parameter space $\Theta\subset\real$ is the interval $(0,\infty)$. For any $\theta>0$ and $i=1,\dots,n$, $Y_i(\theta)$ can be expressed as $\theta U_i$ where $U_i$ is i.i.d. $U(0,1)$. Hence, we have $\widehat{\pi}_n(\theta,\normalfont\text{W}) = \theta \normalfont\text{W}$, where $\normalfont\text{W}$ is a $U_{(n)}$ random variable. Here, $\psi(\theta)=\theta$  and $\vartheta_0=\theta_0$. A perfect matching in~\cref{eq:def:ImpBoot-estimator:bomegas} is possible for all realizations $\normalfont\text{w},\normalfont\text{w}^\ast$ of $\normalfont\text{W}$ and $\normalfont\text{W}^\ast$, respectively; it is given by $\widecheck{\theta}_n(\normalfont\text{w}^\ast|\normalfont\text{w}) = \theta_0 \normalfont\text{w}/\normalfont\text{w}^\ast$. In this example,  for any realization $\normalfont\text{w}$ of $\normalfont\text{W}$ and any $x\in(0,1)$, we have $h_{\normalfont\text{w}}(x)\vcentcolon=\theta_0\normalfont\text{w}x$ and $h(x)\vcentcolon=1/x$. It is straightforward to see that \cref{assum:exact:psi:factor} holds. 
\end{Example}
\begin{Example}[Pareto distribution]\label{example:Pareto}
    Suppose that $Y_1,\dots,Y_n$ are i.i.d. $\text{Pareto}(\mu_0,\alpha_0)$, where the minimum and the shape parameters are $\mu_0>0$ and $\alpha_0>0$, respectively. Here, the parameter space $\bT\subset\real^2$ is the set $(0,\infty)^2$. Take the MLE as the given initial estimator. It can be expressed as $\hbpi_n(\bt,\W)=(\mu\normalfont\text{W}_1^{-\nicefrac{1}{\alpha}},\alpha\normalfont\text{W}_2)^\top$, where $\normalfont\text{W}_1$ is a $U_{(n)}$ random variable and $\normalfont\text{W}_2$ follows an inverse gamma random distribution; $\normalfont\text{W}_1$ is independent of $\normalfont\text{W}_2$~\citep{malik1970estimation}. We obtain
    \begin{equation}\label{eq:pareto:check:theta} \wc\bt_n(\w^\ast|\w)=\left(\begin{array}{c}
    \mu_0\frac{{(\normalfont\text{w}_1^\ast)}^{\normalfont\text{w}_2^\ast/(\alpha_0\normalfont\text{w}_2)}}{\normalfont\text{w}_1^{1/\alpha_0}}\\       \alpha_0\frac{\normalfont\text{w}_2}{\normalfont\text{w}_2^\ast} 
        \end{array}\right),
    \end{equation}
    where $\w=(\normalfont\text{w}_1,\normalfont\text{w}_2)^\top$,$\w^\ast=(\normalfont\text{w}_1^\ast,\normalfont\text{w}_2^\ast)^\top$ and $\w,\w^\ast\in(0,1)\times(0,\infty)$. 
    
    If $\vartheta_0=\psi(\bt_0)\vcentcolon=\alpha_0$, we immediately see that \cref{assum:exact:psi:factor} is satisfied as the functions $h(\W^\ast)\vcentcolon=1/\normalfont\text{w}_2^\ast$ and $h_{\w}(x)\vcentcolon=\alpha_0\normalfont\text{w}_2x$ are defined similarly to the ones in \cref{eg:unif}. 
    
    If $\vartheta_0=\psi(\bt_0)\vcentcolon=\mu_0$, then, using \cref{eq:pareto:check:theta}, we have, 
    \begin{align}\label{eq:pareto:check:vartheta}
    \begin{split}
        \widecheck\vartheta_n(\w^\ast|\w) &= \mu_0\frac{{(\normalfont\text{w}_1^\ast)}^{\normalfont\text{w}_2^\ast/(\alpha_0\normalfont\text{w}_2)}}{\normalfont\text{w}_1^{1/\alpha_0}} = \mu_0 \left\{{(\normalfont\text{w}_1^\ast)}^{\normalfont\text{w}_2^\ast}\normalfont\text{w}_1^{-\normalfont\text{w}_2}\right\}^{1/\alpha_0\normalfont\text{w}_2} \\ 
        &=\vcentcolon\mu_0 \left\{h(\w^\ast)\normalfont\text{w}_1^{-\normalfont\text{w}_2}\right\}^{1/\alpha_0\normalfont\text{w}_2} =\vcentcolon h_{\w}\{h(\w^\ast)\},
    \end{split}
    \end{align}
    where $h(\w^\ast)\in(0,1)$ and $h_\w(x)$ is strictly increasing for all $\w,\w^\ast\in(0,1)\times(0,\infty)$ and all $x\in(0,1)$. Hence, \cref{assum:exact:psi:factor} is satisfied.
\end{Example}
\begin{Example}[\citealp{Andrews:00}]\label{Example:Andrews}
    Suppose that $Y_1,\dots,Y_n$ are i.i.d. $\mathcal{N}(\theta_0,1)$ where $\theta_0\geq0$ and $\widehat\pi_n=\max(n^{-1}\sum_{i=1}^nY_i,0)$. Here, the parameter space $\Theta\subset\real$ is the interval $[0,\infty)$, $\psi(\theta)=\theta$  and $\vartheta_0=\theta_0$. It is shown in \citealp{Andrews:00}, that for $\vartheta_0=0$ the standard parametric and nonparametric bootstrap are inconsistent. The implicit bootstrap distribution based on the same estimator $\widehat\pi_n$, on the other hand, provides exact CIs. Indeed, some simple calculations lead to the following expression for an implicit bootstrap estimate: 
    \begin{equation*}    \widecheck\vartheta_n(\normalfont\text{w},\normalfont\text{w}^\ast)=\max\{\max(\vartheta_0+\normalfont\text{w},0)-\normalfont\text{w}^\ast,0\},
    \end{equation*} 
    where $\normalfont\text{w},\normalfont\text{w}^\ast$ are realizations from $\normalfont\text{W}\sim\mathcal{N}(0,n^{-1})$. For any realization $\normalfont\text{w}$, and any $x\in\real$, we have $h_{\normalfont\text{w}}(x)=\max\{\max(\theta_0+\normalfont\text{w},0)-x,0\}$, i.e., decreasing continuous functions, and $h(\normalfont\text{w}^\ast)=\normalfont\text{w}^\ast$. In this example, the implicit bootstrap estimator has a point mass distribution at 0 with probability $\Pr\{\normalfont\text{W}^\ast\geq\max(\theta_0+\normalfont\text{W},0)\vert\normalfont\text{W}\}\leq1/2.$ Thus, if we consider $\alpha$-level quantile with $\alpha>1/2$, we have $\widecheck\vartheta^\ast_{n,\alpha}>0$. Direct calculations lead to the exact coverage property of the implicit bootstrap CIs: if $\vartheta_0=0$,then for any $\alpha\in(0,1/2]$, $\Pr(\vartheta_0<\widecheck\vartheta^\ast_{n,\alpha})=\alpha$ and for any $\alpha\in(1/2,1)$, $\Pr(\vartheta_0<\widecheck\vartheta^\ast_{n,\alpha})=1\geq\alpha$. Hence, for any $\alpha\in(0,1)$, we have $\Pr(\vartheta_0\leq\widecheck{\vartheta}^\ast_{n, \alpha})\geq\alpha$.
\end{Example}

\section{Simulation studies}\label{sec:simulation}

In this section, we present empirical evidence of the coverage accuracy of the implicit bootstrap in three parametric scenarios where inference is complex. First, we examine a regression model with censored data, specifically a Student's \textit{t} censored linear regression, as detailed in~\cref{sec:simu-censored}. Second, we investigate a heavy-tailed distribution, the Lomax distribution, in the potential presence of outliers, employing a robust estimator as discussed in~\cref{sec:sim-rob}. Third, we explore a queuing model with difficult-to-evaluate likelihood function, as described in~\cref{sec:queuing-models}.

We report the empirical coverages of the upper endpoints of one-sided intervals with $\alpha\in(0.9,1)$. We compare the implicit bootstrap, the percentile parametric bootstrap and the CIs based on the asymptotic theory. When computationally feasible, other bootstrap schemes are also included. Overall, the implicit bootstrap consistently provides the most accurate coverages. Other methods can be highly unreliable, especially with relatively small sample sizes, reinforcing the theoretical assertion that the implicit bootstrap is second-order accurate in complex inferential parametric problems. Furthermore, the possibility to use a simple inconsistent estimator as the initial estimator in all cases demonstrates the practical applicability of the implicit bootstrap across a wide range of parametric models.

\subsection{Student's \textit{t} censored linear regression model}\label{sec:simu-censored}

An extension to the linear regression model involves considering error distributions more general than the normal distribution to accommodate potential heavy tails. This model can also be seen as an approach to provide robust estimators for the regression parameters  \citep[see, e.g.,][]{LaLiTa:89,HeSiWa:00,HeCuSi:04,AzGe:08}. Here, we consider the case of i.i.d. Student's \textit{t} distributed errors \citep[see, e.g.,][]{Emperors1997}. Specifically, the Student's \textit{t} linear regression model for responses $Y_i,i=1,\ldots,n$ and associated covariates $\x_i=[x_{ij}]_{j=0,\ldots,p}$ can be expressed as
\begin{equation}\label{eqn:student-reg}
    Y_i = \x_i^\top\bb+ \sigma\varepsilon_i, \quad i=1,\dots,n,
\end{equation}
where $\bb=(\beta_0\;\beta_1\;\ldots\;\beta_{p})^\top$ and $\varepsilon_i$ are i.i.d. Student's \textit{t} distributed with $\nu$ degrees of freedom.

Moreover, as argued in \citet{arellano2012student}, the response variable might be observed on a limited support, typically only positive values, which can be modeled as a censoring feature. For instance, they considered the case of average hourly earnings (wage rates) of married women recorded alongside a set of covariates. The wage rates were considered as censored for women who did not work in the survey year. Following~\citet{arellano2012student}, we assume that instead of observing $Y_i,i=1,\dots,n,$ we observe $Y^c_i=Y_i\,\mathbb{I}(Y_i>0)$, where $\mathbb{I}(\cdot)$ is the indicator function. Let $n_1\vcentcolon=\sum_{i=1}^n\mathbb{I}(Y_i>0)$ and $n_0\vcentcolon=n-n_1$ be the number of responses and censored responses, respectively. Assuming the censored proportion $n_0/n$ is known along with the covariates for the censored responses, and letting $d_i\vcentcolon=\mathbb{I}(y_i>0)$, the likelihood function is 
\begin{equation*}
    \prod_{i=1}^n\left\{1-S_\nu\left(\frac{\x_i^\top\bb}{\sigma}\right)\right\}^{1-d_i}\left\{\frac{1}{\sigma}s_\nu\left(\frac{y_i-\x_i^\top\bb}{\sigma}\right)\right\}^{d_i},
\end{equation*}
where $S_\nu$ and $s_\nu$ are the Student's \textit{t} distribution and density functions with $\nu$ degrees of freedom. The vector $\bt$ is formed by $\bb$, $\sigma$ and $\nu$. To compute the MLE,~\citet{arellano2012student} propose using the EM algorithm. \citet{garay2017linear} extend the EM algorithm for censored data to other types of distributions (available in the \texttt{R} package \texttt{SMNCensReg}). Inference for the model's parameters can be performed using the asymptotic covariance matrix by plugging in the parameter estimates obtained from \texttt{SMNCensReg}, or by performing the percentile parametric bootstrap. The studentized bootstrap is computationally prohibitive as the EM algorithm to compute the MLE is itself numerically intensive.

In this simulation study, we propose using the implicit bootstrap, with the MLE of the Student's \textit{t} linear regression model in~\cref{eqn:student-reg} (ignoring the censoring mechanism) as the initial estimator. We refer to this initial estimator as the \textit{naive MLE}. Although this is clearly an inconsistent estimator for the Student's \textit{t} censored linear regression model's parameters, it is used to compute the implicit bootstrap estimator. We compare the coverage for one-sided CIs of the implicit bootstrap based on the naive MLE and the percentile parametric bootstrap based on the MLE computed using the EM algorithm. As a benchmark, we also consider the CIs obtained using the estimated asymptotic covariance matrix from \texttt{SMNCensReg}. 

The simulation setting is based on the application on housewives' average hourly earnings modeled in \citet{arellano2012student} and \citet{garay2017linear} using Student's~\textit{t} censored linear regression. The original data, described in~\citet{mroz1987sensitivity}, consists of 753 married white women aged 30 to 60 years in 1975. There are fifteen available covariates, of which four were considered in \citet{arellano2012student} and \citet{garay2017linear}: the housewives' age, years of schooling, number of children younger than six years old and  number of children aged between 6 to 19 years. In the simulation study, we consider the full dataset but we set the slope coefficients to zero for the covariates other than the four selected ones. For these, we set $\beta_1=-1,\beta_2=1.5,\beta_3=-0.5,\beta_4=-1.5$,  $\sigma^2=2$, for the error variance and  $\nu=2$, for the degrees of freedom. The covariates are centered and scaled, so the conditional response expectation is a function of the intercept. We choose three intercept values to vary the percentage of censoring: $\beta_0=2$ ($20\%$  censoring), $\beta_0=4$ ($10\%$ censoring) and $\beta_0=6$ ($5\%$ censoring). We consider three sample sizes: $n=100$, $n=200$ and $n=400$, and randomly subsample from the full design matrix to create the designs for these  sample sizes. For each case, we simulate $10,000$ response variables of size $n$. The bootstrap distributions for all methods are obtained using $1,000$ bootstrap samples.

\begin{figure}[htbp]
    \centering
    \includegraphics[scale=.75]{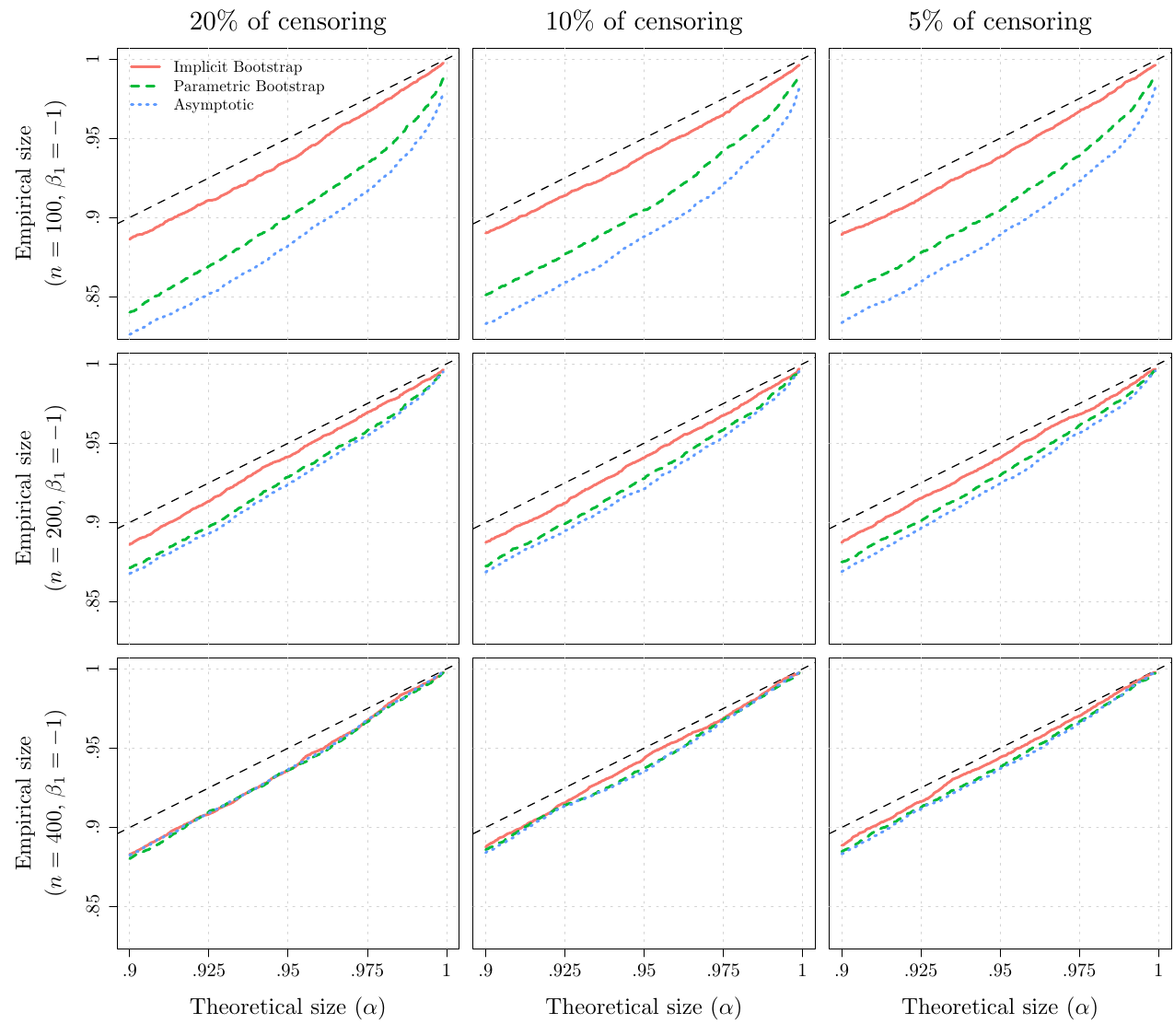}
    \caption{Coverages as a function of the confidence levels $\alpha\in (0.9,1)$ of CIs for $\beta_1=-1.0$. The simulation setting is the Student's \textit{t} censored linear regression model with data left-censored at $0$. The coverages are computed using (in red solid line) the implicit bootstrap based on a naive MLE, and (in green dashed line) the percentile parametric bootstrap and (in blue dotted line) the asymptotic method both based on the EM algorithm. The percentages of censoring are 20\% (first column), 10\% (second column) and 5\% (third column). The sample sizes are $n=100$ (first row), $n=200$ (second row) and $n=400$ (third row). The black dashed line represents the nominal confidence level. We used $10,000$ simulated data and $1,000$ bootstrap samples in all settings.}
    \label{fig:cov-student-reg}
\end{figure}

In \cref{fig:cov-student-reg}, we compare the performance of the implicit bootstrap based on the naive MLE, the percentile parametric bootstrap based on the MLE and the asymptotic method. The CIs are presented for the parameter $\beta_1=-1.0$, with similar results for other parameters.

The censoring proportion does not seem to affect the coverages across methods and sample sizes. For the smaller sample sizes $n=100$ and $n=200$, the percentile parametric bootstrap and the asymptotic method provide quite liberal coverages, with the asymptotic method being the most liberal, showing a larger deviation from the confidence level $\alpha$ compared to the implicit bootstrap. This deviation decreases with larger sample sizes. For instance, considering 10\% of censoring with $n=100$, at the confidence level of 95\%, the coverage for the implicit bootstrap is of 93.9\%, while it is 90.5\% for the percentile parametric bootstrap and 88.8\% for the asymptotic method. With $n=200$, these coverages are of 94.1\% for the implicit bootstrap, 92.8\% for the percentile parametric bootstrap, and 92.1\% for the asymptotic method. We also compare the average CI lengths for two-sided 95\% CIs for $n=400$, since for this sample size the coverages are comparable across methods. These CI lengths are reported in~\cref{table:tobit}. The different methods produce comparable lengths, indicating that there is no loss of efficiency for the implicit bootstrap even though it is based on the naive MLE. The simulation study confirms the theoretical advantages of the implicit bootstrap in terms of coverage accuracy, demonstrating superior performance and accuracy compared to alternative methods. 

\begin{table}
\centering
\begin{tabular}{lccc}
\toprule 
Method & 20\% of censoring & 10\% of censoring & 5\% of censoring \\
\midrule 
 & \multicolumn{3}{c}{Coverages (95\% nominal level)} \\
\cmidrule(lr){2-4}
Implicit Bootstrap & 0.9444 & 0.9423 & 0.9447 \\
Parametric Bootstrap & 0.9384 & 0.9394 & 0.9375 \\
Asymptotic & 0.9361 & 0.9363 & 0.9359 \\
\cmidrule(lr){2-4}
 &\multicolumn{3}{c}{Average CI lengths} \\
\cmidrule(lr){2-4}
Implicit Bootstrap & 0.5866 & 0.5309 & 0.5191 \\
Parametric Bootstrap & 0.5347 & 0.5077 & 0.5023 \\
Asymptotic & 0.5292 & 0.5025 & 0.4975 \\
\bottomrule 
\end{tabular}
\caption{Coverages and average lengths of two-sided 95\% CIs for $\beta_1=-1$. The simulation setting is the Student's \textit{t} linear regression model with left-censored data at $0$. The coverages and lengths are computed using the implicit bootstrap based on a naive MLE, the percentile parametric bootstrap and the asymptotic method both based on the EM algorithm. The percentages of censoring are 20\% (first column), 10\% (second column) and 5\% (third column). The sample size is $n=400$. We used $10,000$ simulated data and $1,000$ bootstrap samples in all settings.}
\label{table:tobit}
\end{table}

\subsection{Heavy-tailed density estimation}\label{sec:sim-rob}

As a heavy-tailed distribution, we consider the widely used Lomax distribution \citep{lomax1954business}, which is frequently applied in fields such as business, economics, actuarial science, queueing theory, internet traffic modeling \citep[see, e.g.,][]{Holland-IEEE-06}, human dynamics \citep[see, e.g.,][]{Barbasi2005}. Also known as the Pareto II distribution~\citep{kleiber2003statistical}, the Lomax distribution is a shifted version of the Pareto distribution, with its support starting at 0. This distribution is particularly useful for fitting upper tails of distributions, where data are sparse, and for deriving quantities associated with extreme events or risk measures \citep[see, e.g.,][]{EmKlMi:97}. Additionally, it is often employed to estimate inequality measures or Lorenz curves \citep[see, e.g.,][]{CoVF:07}. 

The density function of the Lomax distribution is given by:
\begin{equation*}
    \frac{q}{b}\left(1+\frac{y}{b}\right)^{-(q+1)}, \quad y>0,
\end{equation*}
where $b>0$, $q>0$. The value of $q$ determines the existence of moments of this distribution; specifically, for $m\in\mathbb{N}^\ast$, $\mathbb{E}[X^m]$ is defined if $q>m$. In applications, small values of $q$ are common; for example, \cite{Holland-IEEE-06} reported values ranging from $0.5$ to $3.3$. It is known that the MLE for the Lomax parameters can be severely biased \citep[see, e.g.,][and the references therein]{giles2013bias}, especially for $q\leq 2$, where the second moment does not exist, leading to poor coverage. Additionally, there are concerns about  the non-robustness of statistics computed from fitted Pareto distributions used in risk theory for insurance or finance, as well as for inequality measures; see, e.g., \cite{PeWe:01}, \cite{Cowell2002WefRank}, \cite{Braza:03}, \cite{DuFi:04}, \cite{DuVF:06}, \cite{CoVF:07} and \cite{AlTeFi:12}. 

To address these issues, robust estimators such as the weighted MLE (WMLE) proposed by \citet{field1994robust} have been used. For the Lomax distribution, the likelihood score function components are:
\begin{eqnarray*}
    s_b(y;\bt)&=& (q+1)\frac{y}{b^2+b y} - \frac{1}{b}, \\
    s_q(y;\bt)&=& \frac{1}{q}-\log\left(1+\frac{y}{b}\right). \\   
\end{eqnarray*}
Letting $\mathbf{s}(y;\bt)=[s_b(y;\bt)\;\; s_q(y;\bt)]^\top$, the WMLE is obtained by solving:
\begin{equation}\label{eqn:WMLE}
    \sum_{i=1}^n w(y_i;\bt)\mathbf{s}(y_i;\bt)-\mathbf{h}(\bt)=\bm{0},   
\end{equation}
where $w(y_i;\bt)$ is a weight function, such as Huber's weights \citep[see, e.g.,][]{hampel1986robust}, and $\mathbf{h}(\bt)\vcentcolon=\int w(y;\bt)\mathbf{s}(y)\d F_\bt$ is included to give Fisher consistency. For the Lomax distribution, this correction factor lacks a closed-form and requires numerical approximations. Alternatively, \cite{moustaki2006bounded} proposed adjusting a robust estimator for consistency using the principle of indirect inference. We adopt this approach to compute the WMLE.

Since the implicit bootstrap does not require a consistent estimator, we use a \textit{naive WMLE} by setting $\mathbf{h}(\bt)=\0$ in~\eqref{eqn:WMLE} as the initial estimator. We compare the implicit bootstrap with the percentile parametric bootstrap and the asymptotic method both based on the WMLE. Additionally, we compute CIs based on the MLE using the implicit bootstrap, the percentile parametric bootstrap, the studentized parametric bootstrap and the BC$_\text{a}$ parametric bootstrap \citep{efron1987better}. 

The simulation study is based on $10,000$ simulated samples of size $n$ drawn from the Lomax distribution with parameters $b=1$ and $q=1.5$. We consider three sample sizes $n=50$, $n=100$ and $n=200$, using $1,000$ bootstrap samples for each method. We compare the coverage of CIs for the parameters of the Lomax distribution. Additionally, since the Lomax distribution is often used to estimate functions of its parameters, such as the survival function $\vartheta(y)=(1 + y/b)^{-q}$, we also compute the coverage of this function at a fixed value for $y$. The studentized and BC$_\text{a}$ bootstrap methods are not used for this function since they are not transformation-respecting. 

\begin{figure}[htbp]
    \centering
    \includegraphics[scale=.75]{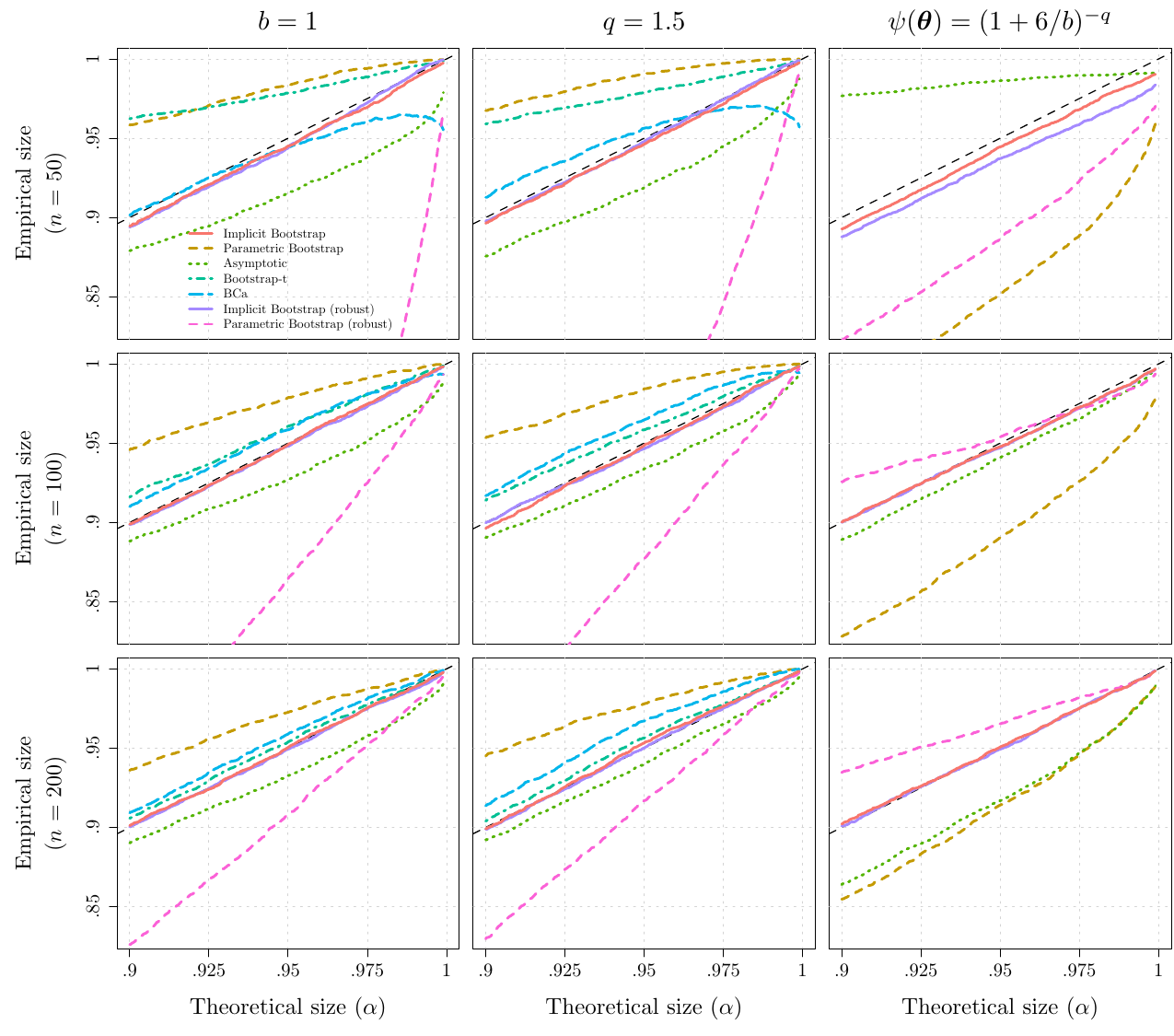}
    \caption{Coverages as a function of the confidence levels $\alpha\in (0.9,1)$ for $b=1$ (first column), $q=1.5$ (second column) and $\vartheta = (1 + 6/b)^{-q}$ (third column). The sample sizes are $n=50$ (first row), $n=100$ (second row) and $n=200$ (third row). Based on the MLE, coverage probabilities are computed using (in red solid line) the implicit bootstrap, (in gold dashed line) the percentile parametric bootstrap, (in green dotted line) the asymptotic method, (in dark green dot-dashed line) the studentized parametric bootstrap and (in blue dashed line) the BC$_\text{a}$ parametric bootstrap. Based on the naive WMLE, the coverages are computed using (in purple solid line) the implicit bootstrap and for the WMLE (in pink dashed line) using the percentile parametric bootstrap. The black dashed line represents the nominal confidence level. We used $10,000$ simulated data and $1,000$ bootstrap samples in all settings.}
    \label{fig:lomax-sim}
\end{figure}

In~\cref{fig:lomax-sim}, we compare the implicit bootstrap, the percentile parametric bootstrap, and, where applicable, the studentized parametric bootstrap and the BC$_\text{a}$ parametric bootstrap, using the coverages obtained by simulations. The last two methods are not used for CIs on the WMLE due to the prohibitive computational time.

The CIs for the Lomax parameters based on the MLE have a seemingly exact coverages only with the implicit bootstrap. Other methods provide inaccurate CIs, especially for $n=50$ and $n=100$, with accuracy improving at $n=200$, where the studentized parametric bootstrap is the most accurate. Comparing the implicit bootstrap based on the naive WMLE and the percentile parametric bootstrap based on the WMLE reveals that the implicit bootstrap provides seemingly exact CIs, whereas the latter produces very liberal coverages. Moreover, in the robust case, our implementation shows that it is on average $1,200$ times faster to compute the implicit bootstrap distribution compared to the percentile parametric bootstrap based on the WMLE.

The comparison of CIs for $\vartheta(y)$ using the implicit bootstrap and the percentile parametric bootstrap based on the MLE leads to a similar conclusion. The latter produces very liberal coverages, while the former yields seemingly exact coverages for $n=100$ and $n=200$, with slightly liberal CIs for $n=50$. The same conclusion applies when comparing the implicit bootstrap based on the naive WMLE and the percentile parametric bootstrap based on the WMLE. These findings highlight the advantages of the implicit bootstrap, not only for its finite sample properties but also for its flexibility as it can be based on an easy-to-obtain inconsistent initial estimator. 

\subsection{Estimation of queueing processes}\label{sec:queuing-models}

While queuing models are widely utilized in management science and engineering, evaluating their likelihood functions is known to be numerically very complex, see, e.g., \citet{asanjarani2021survey}. When only departure times $y_1,y_2,\dots,y_n$ are observed, the challenge arises from the exponentially increasing number of terms that must be integrated with increasing sample size to evaluate the likelihood. The M/G/1 queue model, for example, has been extensively studied, including by~\citet{heggland2004estimating}, who examined a scenario where service times are uniformly distributed over the interval $[\theta_1,\theta_2]=[0.3, 0.9]$ and inter-departure times are exponentially distributed with rate $\theta_3=1$. This represents realistic conditions where the expected service time is moderately shorter than the expected inter-departure time, thus balancing the system to avoid persistent empty queues or constant busyness. They used two sample statistics and the MLE of an auxiliary model with a closed-form likelihood function as an initial estimator, subsequently corrected using indirect inference. Bayesian approaches to this problem have also been explored by~\citet{blum2010non} and~\citet{fearnhead2012constructing}. 

Indirect inference estimators are numerically intensive, making the construction of CIs through the bootstrap prohibitively costly. Moreover, the asymptotic covariance matrix is not always known or easily estimated for all parameters. For this setting, we propose using the MLE of the location-scale Fréchet distribution for independent observations \citep[see, e.g.,][]{smith1985maximum} as the initial estimator. We refer to this estimator as the \textit{naive MLE} in this section. The choice for this initial estimator is motivated by: (i) the Fréchet distribution has the same number of parameters as the M/G/1 queue model considered; (ii) the location parameter of the Fréchet distribution is the minimum of its support and it naturally links to $\theta_1$, the minimum support of the M/G/1 queue model; (iii) the Fréchet distribution can reasonably fit observations from this M/G/1 queue model; (iv) it is simple to compute the MLE of the Fréchet distribution; (v) the MLE of the Fréchet distribution can be expressed as a Z-estimator so we can use a computationally advantageous procedure for solving the implicit bootstrap estimator. The naive MLE is not consistent for the parameters of the M/G/1 queue model and this choice highlights the flexibility of the implicit bootstrap due to its insensitivity to the choice of the initial estimators.

We compute CIs for all parameters $\theta_1,\theta_2,\theta_3$ using the implicit bootstrap based on the naive MLE. For comparison, we also compute CIs using the percentile parametric bootstrap based on the naive MLE corrected for consistency through indirect inference, that we refer to as the indirect inference estimator (IIE). While valid, the parametric bootstrap is computationally intensive, as each bootstrap sample requires computing an IIE. As a benchmark, we also consider the CIs obtained following the asymptotic method using the estimated covariance matrix from a parametric bootstrap. We use $1,000$ bootstrap samples for each method.

\begin{figure}[htbp]
    \centering
    \includegraphics[scale=.75]{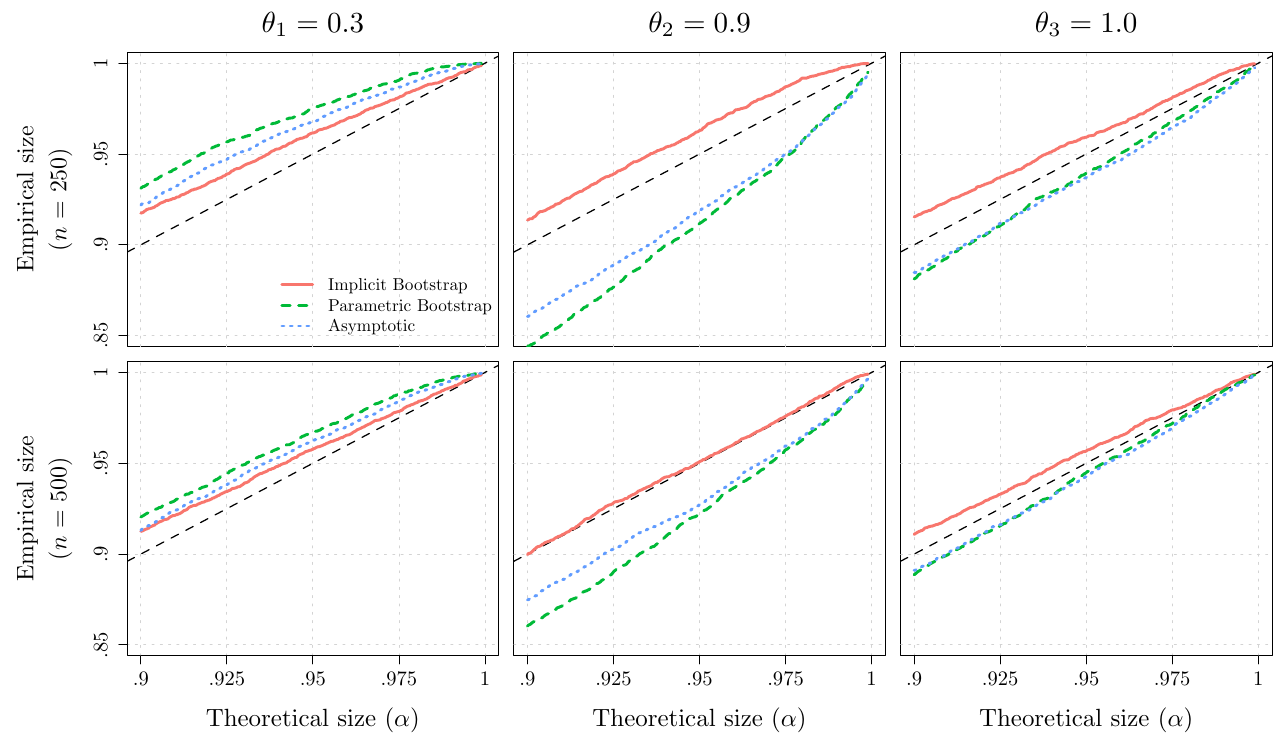}
    \caption{Coverages as a function of the confidence levels $\alpha \in (0.9, 1)$ for $\theta_1 = 0.3$ (first column), $\theta_2 = 0.9$ (second column), and $\theta_3 = 1.0$ (third column). The sample sizes are $n=250$ (first row) and $n=500$ (second row). Coverage probabilities are computed using (in red solid line) the implicit bootstrap  based on the naive MLE, and (in green dashed line) the percentile parametric bootstrap and (in blue dotted line) the asymptotic method both based on the IIE. The black dashed line represents the nominal confidence level. We used $10,000$ simulated data and $1,000$ bootstrap samples in all settings. 
    }
    \label{fig:mg1-sim}
\end{figure}

\cref{fig:mg1-sim} illustrates the empirical coverage probability for CIs of the parameters in the M/G/1 queue model for two sample sizes, $n=250$ and $n=500$. For both sample sizes, the implicit bootstrap consistently demonstrates accurate coverages close to the nominal confidence levels. The percentile parametric bootstrap and the asymptotic method, while generally reliable, exhibit slightly more variability in their coverage probabilities. Notably, for $\theta_2$, the implicit bootstrap CIs have coverage probabilities that are closer to the nominal level compared to the ones of the percentile parametric bootstrap and the asymptotic method. As the sample size increases, all methods improve in accuracy.

The comparison of computational times required to construct CIs using the implicit bootstrap, parametric bootstrap, and obtaining point estimates with indirect inference reveals a clear numerical efficiency advantage for the implicit bootstrap method. For the smaller sample size of $n = 250$, the implicit bootstrap achieves a median completion time of 2.56 seconds, while the parametric bootstrap takes approximately 3.68 hours. The substantial time required for the parametric bootstrap arises from the need to repeat the indirect inference procedure $1,000$ times to obtain the IIE, which, for one point estimate, needs a median completion time of 12.33 seconds, which is more time than deriving the implicit bootstrap distribution for $1,000$ estimates. For the larger sample size of $n = 500$, the implicit bootstrap requires a median time around 4.50 seconds, whereas the parametric bootstrap methods need a median time of approximately 6.75 hours. This significant difference underscores the computational efficiency of the implicit bootstrap, making it a highly practical option for large-scale simulations and real-world applications, especially when computational resources and time are limiting factors.

\section{Conclusion}\label{sec:conclusion}

In this paper, we have introduced the implicit bootstrap as a novel and efficient approach for constructing confidence intervals for target quantities when dealing with asymptotically biased estimators. The salient point of this method is that, being insensitive to the initial estimator inconsistency, it circumvents the need for burdensome bias correction procedures while maintaining statistical accuracy. By avoiding these corrections, the implicit bootstrap is suitable for situations where obtaining a consistent estimator is challenging. Even when a consistent estimator is available, the implicit bootstrap is a worthy addition to parametric bootstrap methods as it is a second-order accurate percentile method that does not require any adjustments.  In practice, this makes our proposed approach particularly useful for large-scale simulations and real-world applications where computational resources and time are limited.

A natural extension of this work includes expanding from parametric models to semi-parametric or non-parametric models, or to consider cases where the dimension of the models is allowed to diverge with the sample size. Such investigation is however left for further research.

\newpage
\bibliographystyle{plainnat.bst}
\bibliography{biblio}

\begin{thebibliography}{70}
\providecommand{\natexlab}[1]{#1}
\providecommand{\url}[1]{\texttt{#1}}
\expandafter\ifx\csname urlstyle\endcsname\relax
  \providecommand{\doi}[1]{doi: #1}\else
  \providecommand{\doi}{doi: \begingroup \urlstyle{rm}\Url}\fi

\bibitem[Alfons et~al.(2012)Alfons, Templ, and Filzmoser]{AlTeFi:12}
A.~Alfons, M.~Templ, and P.~Filzmoser.
\newblock {Robust Estimation of Economic Indicators from Survey Samples Based on Pareto Tail Modelling}.
\newblock \emph{Journal of the Royal Statistical Society Series C: Applied Statistics}, 62:\penalty0 271--286, 2012.

\bibitem[Andrews(1992)]{andrews1992generic}
D.~WK Andrews.
\newblock Generic uniform convergence.
\newblock \emph{Econometric theory}, 8\penalty0 (02):\penalty0 241--257, 1992.

\bibitem[Andrews(2000)]{Andrews:00}
D.~WK Andrews.
\newblock Inconsistency of the bootstrap when a parameter is on the boundary of the parameter space.
\newblock \emph{Econometrica: Journal of the Econometric Society}, 68:\penalty0 399--405, 2000.

\bibitem[Arellano-Valle et~al.(2012)Arellano-Valle, Castro, Gonz{\'a}lez-Far{\'\i}as, and Mu{\~n}oz-Gajardo]{arellano2012student}
R.~B. Arellano-Valle, L.~M. Castro, G.~Gonz{\'a}lez-Far{\'\i}as, and K.~A. Mu{\~n}oz-Gajardo.
\newblock Student-t censored regression model: properties and inference.
\newblock \emph{Statistical Methods \& Applications}, 21:\penalty0 453--473, 2012.

\bibitem[Arvanitis and Demos(2018)]{arvanitis2018validity}
S.~Arvanitis and A.~Demos.
\newblock On the validity of {Edgeworth} expansions and moment approximations for three indirect inference estimators.
\newblock \emph{Journal of Econometric Methods}, 7\penalty0 (1):\penalty0 20150009, 2018.

\bibitem[Asanjarani et~al.(2021)Asanjarani, Nazarathy, and Taylor]{asanjarani2021survey}
A.~Asanjarani, Y.~Nazarathy, and P.~Taylor.
\newblock A survey of parameter and state estimation in queues.
\newblock \emph{Queueing Systems}, 97:\penalty0 39--80, 2021.

\bibitem[Azzalini and Genton(2008)]{AzGe:08}
A.~Azzalini and M.~G. Genton.
\newblock Robust likelihood methods based on the skew-t and related distributions.
\newblock \emph{International Statistical Review / Revue Internationale de Statistique}, 76:\penalty0 106--129, 2008.

\bibitem[Barab\`asi(2005)]{Barbasi2005}
A.~L. Barab\`asi.
\newblock The origin of bursts and heavy tails in human dynamics.
\newblock \emph{Nature}, 435:\penalty0 207--211, 2005.

\bibitem[Bhattacharya and Ghosh(1978)]{bhattacharya1978validity}
R.~N. Bhattacharya and J.~K. Ghosh.
\newblock On the validity of the formal {Edgeworth} expansion.
\newblock \emph{Annals of Statistics}, 6\penalty0 (2):\penalty0 434--451, 1978.

\bibitem[Bhattacharya and Rao(1976)]{bhattacharya1976normal}
R.~N. Bhattacharya and R.~R. Rao.
\newblock \emph{Normal approximation and asymptotic expansions}.
\newblock SIAM, 1976.

\bibitem[Bickel and Freedman(1981)]{bickel1981some}
P.~J. Bickel and D.~A. Freedman.
\newblock Some asymptotic theory for the bootstrap.
\newblock \emph{The annals of statistics}, pages 1196--1217, 1981.

\bibitem[Blum and Fran{\c{c}}ois(2010)]{blum2010non}
M.~G.~B. Blum and O.~Fran{\c{c}}ois.
\newblock Non-linear regression models for approximate bayesian computation.
\newblock \emph{Statistics and Computing}, 20\penalty0 (1):\penalty0 63--73, 2010.

\bibitem[Brazauskas(2003)]{Braza:03}
V.~Brazauskas.
\newblock Influence functions of empirical nonparametric estimators of net insurance premiums.
\newblock \emph{Insurance: Mathematics and Economics}, 32:\penalty0 115--133, 2003.

\bibitem[Breusch et~al.(1997)Breusch, Robertson, and Welsh]{Emperors1997}
T.~S. Breusch, J.~C. Robertson, and A.~H. Welsh.
\newblock The emperor's new clothes: a critique of the multivariate t regression model.
\newblock \emph{Statistica Neerlandica}, 51:\penalty0 269--286, 1997.

\bibitem[Cavaliere et~al.(2024)Cavaliere, Gon{\c{c}}alves, Nielsen, and Zanelli]{cavaliere2024bootstrap}
G.~Cavaliere, S.~Gon{\c{c}}alves, M.~O. Nielsen, and E.~Zanelli.
\newblock Bootstrap inference in the presence of bias.
\newblock \emph{Journal of the American Statistical Association}, pages 1--11, 2024.

\bibitem[Chambers(1967)]{chambers1967methods}
J.~M. Chambers.
\newblock On methods of asymptotic approximation for multivariate distributions.
\newblock \emph{Biometrika}, 54\penalty0 (3-4):\penalty0 367--383, 1967.

\bibitem[Chibisov(1974)]{chibisov1974asymptotic}
D.~M. Chibisov.
\newblock An asymptotic expansion for a class of estimators containing maximum likelihood estimators.
\newblock \emph{Theory of Probability \& Its Applications}, 18\penalty0 (2):\penalty0 295--303, 1974.

\bibitem[Cowell and Victoria-Feser(2002)]{Cowell2002WefRank}
F.~A. Cowell and M.-P. Victoria-Feser.
\newblock Welfare rankings in the presence of contaminated data.
\newblock \emph{Econometrica: Journal of the Econometric Society}, 70:\penalty0 1221--1233, 2002.

\bibitem[Cowell and Victoria-Feser(2007)]{CoVF:07}
F.~A Cowell and M.-P. Victoria-Feser.
\newblock Robust stochastic dominance: A semi-parametric approach.
\newblock \emph{The Journal of Economic Inequality}, 5:\penalty0 21--37, 2007.

\bibitem[Cristea(2017)]{cristea2017global}
M.~Cristea.
\newblock On global implicit function theorem.
\newblock \emph{Journal of Mathematical Analysis and Applications}, 456\penalty0 (2):\penalty0 1290--1302, 2017.

\bibitem[DiCiccio and Romano(1995)]{diciccio1995bootstrap}
T.~J. DiCiccio and J.~P. Romano.
\newblock On bootstrap procedures for second-order accurate confidence limits in parametric models.
\newblock \emph{Statistica Sinica}, pages 141--160, 1995.

\bibitem[Dupuis and Field(2004)]{DuFi:04}
D.~J. Dupuis and C.~Field.
\newblock Large wind speeds: Modeling and outlier detection.
\newblock \emph{Journal of Agricultural, Biological and Environmental Sciences}, 9:\penalty0 105--121, 2004.

\bibitem[Dupuis and Victoria-Feser(2006)]{DuVF:06}
D.~J. Dupuis and M.-P. Victoria-Feser.
\newblock A robust prediction error criterion for pareto modeling of upper tails.
\newblock \emph{Canadian journal of statistics}, 34:\penalty0 639--358, 2006.

\bibitem[Efron(1979)]{efron1979bootstrap}
B.~Efron.
\newblock Bootstrap methods: Another look at the jackknife.
\newblock \emph{The Annals of Statistics}, 7\penalty0 (1):\penalty0 1--26, 1979.

\bibitem[Efron(1987)]{efron1987better}
B.~Efron.
\newblock Better bootstrap confidence intervals.
\newblock \emph{Journal of the American statistical Association}, 82\penalty0 (397):\penalty0 171--185, 1987.

\bibitem[Efron and Tibshirani(1994)]{efron1994introduction}
B.~Efron and R.~J. Tibshirani.
\newblock \emph{An introduction to the bootstrap}.
\newblock CRC press, 1994.

\bibitem[Embrechts et~al.(1997)Embrechts, Kl\"uppelberg, and Mikosch]{EmKlMi:97}
P.~Embrechts, C.~Kl\"uppelberg, and T.~Mikosch.
\newblock \emph{Modelling Extremal Events for Insurance and Finance}.
\newblock Springer, Berlin, 1997.

\bibitem[Fearnhead and Prangle(2012)]{fearnhead2012constructing}
P.~Fearnhead and D.~Prangle.
\newblock Constructing summary statistics for approximate bayesian computation: semi-automatic approximate bayesian computation.
\newblock \emph{Journal of the Royal Statistical Society: Series B}, 74\penalty0 (3):\penalty0 419--474, 2012.

\bibitem[Field and Smith(1994)]{field1994robust}
C.~Field and B.~Smith.
\newblock Robust estimation - a weighted maximum likelihood approach.
\newblock \emph{International Statistical Review}, 62:\penalty0 405--424, 1994.

\bibitem[Gallant and Tauchen(1996)]{gallant1996moments}
A.~R. Gallant and G.~Tauchen.
\newblock Which moments to match?
\newblock \emph{Econometric Theory}, 12\penalty0 (4):\penalty0 657--681, 1996.

\bibitem[Garay et~al.(2017)Garay, Lachos, Bolfarine, and Cabral]{garay2017linear}
A.~M. Garay, V.~H. Lachos, H.~Bolfarine, and C.~RB Cabral.
\newblock Linear censored regression models with scale mixtures of normal distributions.
\newblock \emph{Statistical Papers}, 58:\penalty0 247--278, 2017.

\bibitem[Giles et~al.(2013)Giles, Feng, and Godwin]{giles2013bias}
D.~E. Giles, H.~Feng, and R.~T. Godwin.
\newblock On the bias of the maximum likelihood estimator for the two-parameter {Lomax} distribution.
\newblock \emph{Communications in Statistics-Theory and Methods}, 42\penalty0 (11):\penalty0 1934--1950, 2013.

\bibitem[Gouri{\'e}roux et~al.(1993)Gouri{\'e}roux, Monfort, and Renault]{gourieroux1993indirect}
C.~Gouri{\'e}roux, A.~Monfort, and E.~Renault.
\newblock Indirect inference.
\newblock \emph{Journal of applied econometrics}, 8\penalty0 (S1), 1993.

\bibitem[Greene(1981)]{greene1981asymptotic}
W.~H. Greene.
\newblock On the asymptotic bias of the ordinary least squares estimator of the tobit model.
\newblock \emph{Econometrica: Journal of the Econometric Society}, pages 505--513, 1981.

\bibitem[Hall(1992)]{hall1992boot}
P.~Hall.
\newblock \emph{The Bootstrap and Edgeworth Expansion}.
\newblock Springer-Verlag, New York, 1992.

\bibitem[Hampel et~al.(1986)Hampel, Ronchetti, Rousseeuw, and Stahel]{hampel1986robust}
F.~R. Hampel, E.~Ronchetti, P.~J. Rousseeuw, and W.~A. Stahel.
\newblock \emph{Robust statistics: the approach based on influence functions}.
\newblock Wiley, 1986.

\bibitem[Hannig(2009)]{hannig2009generalized}
J.~Hannig.
\newblock On generalized fiducial inference.
\newblock \emph{Statistica Sinica}, pages 491--544, 2009.

\bibitem[Hannig et~al.(2016)Hannig, Iyer, Lai, and Lee]{hannig2016:GFIreview}
J.~Hannig, H.~Iyer, R.~C.~S. Lai, and T.~C.~M. Lee.
\newblock Generalized fiducial inference: A review and new results.
\newblock \emph{Journal of the American Statistical Association}, pages 1346--1361, 2016.

\bibitem[He and Cui(2004)]{HeCuSi:04}
X.~He and D.~Cui, H.~Simpson.
\newblock Longitudinal data analysis using t-type regression.
\newblock \emph{Journal Statistical Planning and Inference}, 122:\penalty0 253--269, 2004.

\bibitem[He et~al.(2000)He, Simpson, and Wang]{HeSiWa:00}
X.~He, D.~Simpson, and G.~Wang.
\newblock Breakdown points of t-type regression estimators.
\newblock \emph{Biometrika}, 87:\penalty0 675--687, 2000.

\bibitem[Heggland and Frigessi(2004)]{heggland2004estimating}
K.~Heggland and A.~Frigessi.
\newblock Estimating functions in indirect inference.
\newblock \emph{Journal of the Royal Statistical Society: Series B}, 66\penalty0 (2):\penalty0 447--462, 2004.

\bibitem[Holland et~al.(2006)Holland, Golaup, and Aghvami]{Holland-IEEE-06}
O.~Holland, A.~Golaup, and A.H. Aghvami.
\newblock Traffic characteristics of aggregated module downloads for mobile terminal reconfiguration.
\newblock \emph{IEE Proceedings - Communications}, 153:\penalty0 683--690, 2006.

\bibitem[Idczak(2016)]{idczak2016generalization}
D.~Idczak.
\newblock On a generalization of a global implicit function theorem.
\newblock \emph{Advanced Nonlinear Studies}, 16\penalty0 (1):\penalty0 87--94, 2016.

\bibitem[Kasy(2019)]{kasy2019uniformity}
M.~Kasy.
\newblock Uniformity and the delta method.
\newblock \emph{Journal of Econometric Methods}, 8\penalty0 (1):\penalty0 20180001, 2019.

\bibitem[Kleiber and Kotz(2003)]{kleiber2003statistical}
C.~Kleiber and S.~Kotz.
\newblock \emph{Statistical Size Distributions in Economics and Actuarial Sciences}.
\newblock Wiley series in probability and statistics. Wiley, New York, 2003.

\bibitem[Komunjer(2012)]{komunjer2012global}
I.~Komunjer.
\newblock Global identification in nonlinear models with moment restrictions.
\newblock \emph{Econometric Theory}, 28\penalty0 (4):\penalty0 719--729, 2012.

\bibitem[Lahiri(2010)]{lahiri2010edgeworth}
S.~N. Lahiri.
\newblock Edgeworth expansions for studentized statistics under weak dependence.
\newblock \emph{The Annals of Statistics}, 38\penalty0 (1):\penalty0 388--434, 2010.

\bibitem[Lange et~al.(1989)Lange, Little, and Taylor]{LaLiTa:89}
K.~L. Lange, R.~J.~A. Little, and J.~M.~G. Taylor.
\newblock Robust statistical modeling using the t distribution.
\newblock \emph{Journal of the American Statistical Association}, 84:\penalty0 881--896, 1989.

\bibitem[Lieberman et~al.(2003)Lieberman, Rousseau, and Zucker]{lieberman2003valid}
O.~Lieberman, J.~Rousseau, and D.~M. Zucker.
\newblock Valid asymptotic expansions for the maximum likelihood estimator of the parameter of a stationary, gaussian, strongly dependent process.
\newblock \emph{The Annals of Statistics}, 31\penalty0 (2):\penalty0 586--612, 2003.

\bibitem[Lomax(1954)]{lomax1954business}
K.~S. Lomax.
\newblock Business failures: Another example of the analysis of failure data.
\newblock \emph{Journal of the American Statistical Association}, 49\penalty0 (268):\penalty0 847--852, 1954.

\bibitem[Malik(1970)]{malik1970estimation}
H.~J. Malik.
\newblock Estimation of the parameters of the {Pareto} distribution.
\newblock \emph{Metrika}, 15\penalty0 (1):\penalty0 126--132, 1970.

\bibitem[Moustaki and Victoria-Feser(2006)]{moustaki2006bounded}
I.~Moustaki and M.-P. Victoria-Feser.
\newblock Bounded-influence robust estimation in generalized linear latent variable models.
\newblock \emph{Journal of the American Statistical Association}, 101\penalty0 (474):\penalty0 644--653, 2006.

\bibitem[Mroz(1987)]{mroz1987sensitivity}
T.~A. Mroz.
\newblock The sensitivity of an empirical model of married women's hours of work to economic and statistical assumptions.
\newblock \emph{Econometrica: Journal of the Econometric Society}, 55\penalty0 (4):\penalty0 765--799, 1987.

\bibitem[Newey(1991)]{newey1991uniform}
W.~K. Newey.
\newblock Uniform convergence in probability and stochastic equicontinuity.
\newblock \emph{Econometrica: Journal of the Econometric Society}, pages 1161--1167, 1991.

\bibitem[Newey and McFadden(1994)]{newey1994large}
W.~K. Newey and D.~McFadden.
\newblock Large sample estimation and hypothesis testing.
\newblock \emph{Handbook of econometrics}, 4:\penalty0 2111--2245, 1994.

\bibitem[Peng and Welsh(2001)]{PeWe:01}
L.~Peng and A.~H. Welsh.
\newblock Robust estimation of the generalized pareto distribution.
\newblock \emph{Extremes}, 4:\penalty0 53--65, 2001.

\bibitem[Petrov(1975)]{petrov1975sums}
V.~V. Petrov.
\newblock \emph{Sums of Independent Random Variables}.
\newblock Springer-Verlag, 1975.

\bibitem[Phillips(1977)]{phillips1977general}
P.~CB Phillips.
\newblock A general theorem in the theory of asymptotic expansions as approximations to the finite sample distributions of econometric estimators.
\newblock \emph{Econometrica: Journal of the Econometric Society}, pages 1517--1534, 1977.

\bibitem[Phillips(2012)]{phillips2012folklore}
P.~CB Phillips.
\newblock Folklore theorems, implicit maps, and indirect inference.
\newblock \emph{Econometrica: Journal of the Econometric Society}, 80\penalty0 (1):\penalty0 425--454, 2012.

\bibitem[Sandberg(1981)]{sandberg1981global}
I.~Sandberg.
\newblock Global implicit function theorems.
\newblock \emph{IEEE Transactions on Circuits and Systems}, 28\penalty0 (2):\penalty0 145--149, 1981.

\bibitem[Sargan(1976)]{sargan1976econometric}
J.~D. Sargan.
\newblock Econometric estimators and the {Edgeworth} approximation.
\newblock \emph{Econometrica: Journal of the Econometric Society}, pages 421--448, 1976.

\bibitem[Skovgaard(1981{\natexlab{a}})]{skovgaard1981edgeworth}
I.~M. Skovgaard.
\newblock Edgeworth expansions of the distributions of maximum likelihood estimators in the general (non iid) case.
\newblock \emph{Scandinavian Journal of Statistics}, pages 227--236, 1981{\natexlab{a}}.

\bibitem[Skovgaard(1981{\natexlab{b}})]{skovgaard1981transformation}
I.~M. Skovgaard.
\newblock Transformation of an {Edgeworth} expansion by a sequence of smooth functions.
\newblock \emph{Scandinavian Journal of Statistics}, pages 207--217, 1981{\natexlab{b}}.

\bibitem[Smith(1993)]{smith1993estimating}
A.~A. Smith.
\newblock Estimating nonlinear time-series models using simulated vector autoregressions.
\newblock \emph{Journal of Applied Econometrics}, 8\penalty0 (S1), 1993.

\bibitem[Smith(1985)]{smith1985maximum}
R.~L. Smith.
\newblock Maximum likelihood estimation in a class of nonregular cases.
\newblock \emph{Biometrika}, 72\penalty0 (1):\penalty0 67--90, 1985.

\bibitem[Thornton and Xie(2024)]{BFF-bridge:2024}
S.~Thornton and M.~Xie.
\newblock Bridging {Bayesian}, {Frequentist} and {Fiducial} {(BFF)} inferences using confidence distribution.
\newblock In J.~Berger, X.-L. Meng, N.~Reid, and M.~Xie, editors, \emph{Handbook on Bayesian, Fiducial and Frequentist (BFF) Inferences}. Chapman \& Hall, 2024.

\bibitem[Van~der Vaart(1998)]{van1998asymptotic}
A.~W. Van~der Vaart.
\newblock \emph{Asymptotic statistics}, volume~3.
\newblock Cambridge university press, 1998.

\bibitem[Van Der~Vaart and Wellner(1996)]{van1996weak}
A.~W. Van Der~Vaart and J.~A. Wellner.
\newblock Weak convergence.
\newblock In \emph{Weak convergence and empirical processes}, pages 16--28. Springer, 1996.

\bibitem[White(1982)]{white1982maximum}
H.~White.
\newblock Maximum likelihood estimation of misspecified models.
\newblock \emph{Econometrica: Journal of the econometric society}, pages 1--25, 1982.

\bibitem[Zhang et~al.(2023)Zhang, Ma, Orso, Karemera, Victoria-Feser, and Guerrier]{zhang2023just}
Y.~Zhang, Y.~Ma, S.~Orso, M.~Karemera, M.-P. Victoria-Feser, and S.~Guerrier.
\newblock Just identified indirect inference estimator: Accurate inference through bias correction.
\newblock \emph{arXiv:2204.07907}, 2023.

\end{thebibliography}

\end{document}